\documentclass[nojss]{jss}
\usepackage{amsmath,amssymb,amsfonts,multirow,longtable,tikz,thumbpdf,lmodern}
\usetikzlibrary{fit,backgrounds,arrows,calc,arrows.meta}
\tikzset{%
  >={Latex[width=2mm,length=2mm]},
  base/.style = {rectangle, rounded corners, draw=black,
                 minimum width=3.2cm, minimum height=0.8cm,
                 text centered, font=\ttfamily},
  input/.style = {base, minimum width=2.2cm, fill=blue!30},
  processing/.style = {base, fill=yellow!30},
  estimation/.style = {base, fill=red!30},
  stats/.style = {base, fill=blue!30},
  output/.style = {base, minimum width=2.2cm, fill=orange!15},
  outer/.style={draw=gray, dashed, fill=green!1, thick, inner sep=5pt, rounded corners},
}

\definecolor{rb1}{HTML}{E495A5}
\definecolor{rb2}{HTML}{BDAB66}
\definecolor{rb3}{HTML}{65BC8C}
\definecolor{rb4}{HTML}{55B8D0}
\definecolor{rb5}{HTML}{C29DDE}

\newcommand{\dquote}[1]{``{#1}''}
\newcommand{\fct}[1]{{\texttt{#1()}\index{#1@\texttt{#1()}}}}
\newcommand{\class}[1]{\dquote{\texttt{#1}}}

\graphicspath{{Figures/}}

\author{Nikolaus Umlauf\\Universit\"at Innsbruck \And
        Nadja Klein\\Humboldt Universit\"at\\zu Berlin \And
        Thorsten Simon\\Universit\"at Innsbruck \And
        Achim Zeileis\\Universit\"at Innsbruck}
\Plainauthor{Nikolaus Umlauf, Nadja Klein, Thorsten Simon, Achim Zeileis}

\title{\pkg{bamlss}: A Lego Toolbox for Flexible {B}ayesian Regression (and Beyond)}
\Plaintitle{bamlss: A Lego Toolbox for Flexible Bayesian Regression (and Beyond)}
\Shorttitle{\pkg{bamlss}: A Lego Toolbox for Flexible Bayesian Regression}
	
\Plainkeywords{GAMLSS, distributional regression, probabilistic forecasting, backfitting, gradient boosting, lasso, MCMC, R}

\Abstract{
  Over the last decades, the challenges in applied regression and in predictive modeling have
  been changing considerably: (1)~More flexible model specifications are needed as big(ger) data
  become available, facilitated by more powerful computing infrastructure. (2)~Full probabilistic modeling
  rather than predicting just means or expectations is crucial in many applications.
  (3)~Interest in Bayesian inference has been increasing both as an appealing framework for
  regularizing or penalizing model estimation as well as a natural alternative to classical
  frequentist inference. However, while there has been a lot of research in all three areas,
  also leading to associated software packages, a modular software implementation that
  allows to easily combine all three aspects has not yet been available. For filling this gap,
  the \proglang{R} package \pkg{bamlss} is introduced for Bayesian additive models for location,
  scale, and shape (and beyond). At the core of the package are algorithms for highly-efficient Bayesian
  estimation and inference that can be applied to generalized additive models (GAMs) or
  generalized additive models for location, scale, and shape (GAMLSS), also known as
  distributional regression. However, its building blocks are designed as ``Lego bricks''
  encompassing various
  distributions (exponential family, Cox, joint models, \dots),
  regression terms (linear, splines, random effects, tensor products, spatial fields, \dots), and
  estimators (MCMC, backfitting, gradient boosting, lasso, \dots).
  It is demonstrated how these can be easily recombined to make classical models more
  flexible or create new custom models for specific modeling challenges.
}

\Address{
  Nikolaus Umlauf, Achim Zeileis, Thorsten Simon\\
  Department of Statistics\\
  Faculty of Economics and Statistics\\
  Universit\"at Innsbruck\\
  Universit\"atsstr.~15\\
  6020 Innsbruck, Austria\\
  E-mail: \email{Nikolaus.Umlauf@uibk.ac.at},\\
  \phantom{E-mail: }\email{Achim.Zeileis@R-project.org}\\
  \phantom{E-mail: }\email{Thorsten.Simon@uibk.ac.at}\\
  URL: \url{http://eeecon.uibk.ac.at/~umlauf/},\\
  \phantom{URL: }\url{http://eeecon.uibk.ac.at/~zeileis/}\\

  Nadja Klein\\
  Humboldt Universit\"at zu Berlin\\
  School of Business and Economics\\
  Applied Statistics\\
  Unter den Linden 6\\
  10099 Berlin, Germany\\
  E-mail: \email{nadja.klein@hu-berlin.de}\\
  URL: \url{https://hu.berlin/NK}
}

\begin{document}

\section{Introduction}

Many modern modeling tasks necessitate flexible regression tools that can deal with:
(1)~Big data sets that can be both long (many observations) and/or wide (many variables or complex effect types).
(2)~Probabilistic forecasts that capture the entire distribution and not only its mean or expectation.
(3)~Enhanced inference infrastructure, typically Bayesian, beyond classical frequentist significance tests.
A popular framework to combine flexible regression with probabilistic modeling
are generalized additive models \citep[GAMs,][]{bamlss:Hastie+Tibshirani:1990}, later extended
to generalized additive models for location, scale, and shape \citep[GAMLSS,][]{bamlss:Rigby+Stasinopoulos:2005},
also known as distributional regression \citep{bamlss:Klein+Kneib+Lang+Sohn:2015} which
encompasses basic (generalized) linear models \citep{bamlss:Nelder+Wedderburn:1972} as special cases.
These regression approaches can also be combined with Bayesian inference
\citep[see e.g.,][]{bamlss:Fahrmeir+Kneib+Lang+Marx:2013} as a natural framework
for penalizing flexible model terms and to overcome potential problems with $p$~values
and classical null hypothesis significance testing \citep{bamlss:ASA:2016}. However,
when fitting such models to big data -- long and/or wide -- classical estimation
techniques using standard algorithms like iteratively weighted least squares (IWLS) or
Markov chain Monte Carlo (MCMC) might not be feasible. Instead, regularized estimation
techniques such as lasso or boosting \citep{bamlss:Friedmann+Hastie+Tibshirani:2010,bamlss:Mayr+Fenske+Hofner:2012}
might be necessary or custom algorithms \citep{bamlss:Wood:2017}. Hence,
to facilitate combining all three aspects discussed above with different estimation
techniques and fitting algorithms, the \pkg{bamlss} package for the \proglang{R}
system for statistical computing \citep{bamlss:R} implements a modular ``Lego toolbox'',
extending the work of \citet{bamlss:Umlauf+Klein+Zeileis:2018}. In this framework
not only the response distribution or the regression terms are ``Lego bricks''
but also the estimation algorithm or the MCMC sampler.

In terms of software infrastructure, the \proglang{R} ecosystem already provides a
rich variety of packages that combine several -- but not all -- of the aspects discussed above.
\begin{itemize}

\item \emph{GAMs and GAMLSSs}
are available in a number of packages, most notably the \pkg{mgcv} package \citep{bamlss:Wood:2017}
and also the \pkg{gamlss} family of packages \citep{bamlss:Stasinopoulos:Rigby:2007}
and \pkg{VGAM} \citep{bamlss:Yee:2009}. The latter two are notable for their support
of a wide range of response distributions. However, for complex predictor structures
and response distributions beyond the exponential family, estimation may be challenging
or subject to numerical instabilities. In contrast, \pkg{mgcv} excels at providing
highly-optimized algorithms for general smooth models \citep{bamlss:Wood+Pya+Saefken:2016}
as well as the dedicated \fct{bam} function for big data that is long and/or wide
\citep{bamlss:Wood+Li+Shaddick:2017}.

\item \emph{Bayesian inference} is not only an increasingly popular alternative to classical
frequentist inference, it is also particularly attractive for hierarchical or multilevel
models and for penalizing regression effects through suitable prior distributions. Also, fully
Bayesian approaches using MCMC are appealing in flexible regression models
for obtaining credible intervals from the posterior samples etc. The \pkg{brms} package
\citep{bamlss:Buerkner:2017} is notable for providing a standard \proglang{R} workflow
for estimating Bayesian multilevel models using \pkg{Stan} \citep{bamlss:stan}.
Also, the above-mentioned \pkg{mgcv} package supports estimation of Bayesian GAMs via its
\fct{jagam} function \citep{bamlss:Wood:2016b} based on \pkg{JAGS} \citep{bamlss:Plummer:2013}.

For more flexibility, going beyond these capabilities, it is in principle possible to 
directly implement custom models using general purpose MCMC software like
\pkg{JAGS}, \pkg{Stan}, or \pkg{WinBUGS} \citep{bamlss:Lunn+Thomas+Best+Spiegelhalter:2000}.
However, for complex models -- e.g., using large data sets, spatial effects, or higher-order interactions --
sampling times from these generic MCMC engines can become long, sometimes prohibitively so.
This has been addressed by dedicated packages for Bayesian additive models, e.g., with the standalone
package \pkg{BayesX}~\citep{bamlss:Brezger+Kneib+Lang:2005} being the first to provide highly-efficient
sampling schemes for very large data sets as well as spatial/multilevel models. An \proglang{R}
interface is available in \pkg{R2BayesX} \citep{bamlss:Umlauf:2015}. Instead of fully Bayesian MCMC
it is also possible to employ posterior mean estimation via the integrated nested Laplace
approximation to estimate flexible Bayesian regression models. This is provided
in the comprehensive \proglang{R} package \pkg{INLA} \citep{bamlss:Rue+Martino+Chopin:2009},
popular for estimating complex spatial Bayesian regression models
\citep[see e.g.,][]{bamlss:Lindgren+Rue:2015,bamlss:Bivand+GomezRubio+Rue:2015}.

\item \emph{Regularized estimation} might be necessary, though, for going beyond the models
described above, especially for large/wide data with many potential regressors and corresponding
effects/interactions/etc. Widely-used approaches for this include the lasso, e.g., as available
for GLM-type models in the \proglang{R} package \pkg{glmnet} \citep{bamlss:Friedmann+Hastie+Tibshirani:2010},
or gradient boosting as available for GAMLSS-type models in the \proglang{R} package
\pkg{gamboostLSS} \citep{bamlss:Hofner+Mayr+Schmid:2014}. However, obtaining MCMC samples from
the posterior distributions corresponding to such models is not easily available in these
packages.

\end{itemize}
In summary, the discussion above highlights that many different packages with different strengths
are already available in \proglang{R}. However, a package combining all the aspects above in a single
framework is not readily available as there are typically limitations with respect to the 
inferential framework, the distributions and/or complexity of the models supported, or the
estimation techniques and fitting algorithms. The package \pkg{bamlss}, available from the Comprehensive
\proglang{R} Archive Network at \url{https://CRAN.R-project.org/package=bamlss}, tries to 
fill this gap with a modular ``Lego'' approach to flexible Bayesian regression providing:
\begin{itemize}
\item The usual \proglang{R} ``look \& feel'' for regression modeling.
\item Estimation of classic (GAM-type) regression models (Bayesian or frequentist).
\item Estimation of flexible (GAMLSS-type) distributional regression models.
\item An extensible ``plug \& play'' approach for regression terms.
\item Modular combinations of fitting algorithms and samplers.
\end{itemize}
Especially the last item is notable because the models in \pkg{bamlss} are not limited to
a specific estimation algorithm but different engines can be plugged in without necessitating
changes in other aspects of the model specification (such as response distributions or
regression terms). By default \pkg{bamlss} is using IWLS-based backfitting for optimizing
the model and IWLS-based MCMC for sampling from the posterior distribution. However,
alternative optimizers and samplers are also implemented that support lasso
or boosting etc. Moreover, the package builds on the well-established \pkg{mgcv} infrastructure
for smooth model terms, uses \proglang{R}'s formula syntax for model specification, and provides
standard extractor methods like \fct{summary}, \fct{plot}, \fct{predict}, etc.

The remainder of this paper is as follows. In Section~\ref{sec:motivation},
three motivating examples illustrate the first steps using \pkg{bamlss} and show cases the flexibility
of the provided infrastructures. Section~\ref{sec:flexreg}
introduces the flexible regression framework in more detail. A thorough introduction of the
\proglang{R} package \pkg{bamlss}, describing the most important building blocks for developing
families, model terms and estimation algorithms, is then given in Section~\ref{sec:package}. In
Section~\ref{sec:application} we highlight the unified modeling approach using a complex
distributional regression model for lighting counts in complex terrain.
Further details and examples about the \pkg{bamlss} package can be found online at
\url{http://www.bamlss.org/}.

\section{Motivating examples} \label{sec:motivation}

This section gives a first quick overview of the functionality of the package. The first
example demonstrates that the usual ``look \& feel'' when using well-established model fitting 
functions like \fct{glm} is an elementary part of \pkg{bamlss}, i.e., first steps and
basic handling of the package should be relatively simple. The second example shows that the
package can deal with a variety of different model terms and that model fitting functions
can easily be exchanged, here, we exemplify this feature by applying a lasso-type estimation engine.
The third example then explains how full distributional regression models can be estimated and show
cases once more the flexibility of the provided modeling infrastructures.

\subsection{Basic Bayesian regression: Logit model} \label{sec:logitmodel}

This example data is taken from the \pkg{AER} package \citep{bamlss:Kleiber+Zeileis:2008} and is about
labor force participation (yes/no) of women in Switzerland 1981 \citep{bamlss:Gerfin:1996}.
The \pkg{bamlss} package and the data can be loaded with
\begin{Schunk}
\begin{Sinput}
R> library("bamlss")
R> data("SwissLabor", package = "AER")
\end{Sinput}
\end{Schunk}
The data frame contains 872 observations of 6 variables, where some of them might have
a nonlinear influence on the response labor \code{participation}. Now, a standard Bayesian 
binomial logit model using the default MCMC algorithm can be fitted. First, the model formula
specified with
\begin{Schunk}
\begin{Sinput}
R> f <- participation ~ income + age + education +
+    youngkids + oldkids + foreign + I(age^2)
\end{Sinput}
\end{Schunk}
Then, to reproduce the results the seed of the random number generator is set to
\begin{Schunk}
\begin{Sinput}
R> set.seed(123)
\end{Sinput}
\end{Schunk}
The model is estimated by
\begin{Schunk}
\begin{Sinput}
R> b <- bamlss(f, family = "binomial", data = SwissLabor)
\end{Sinput}
\end{Schunk}
Note that the default number of iterations for the MCMC sampler is 1200, the burnin-phase is 200 and
thinning is 1. The reason is that during the modeling process, users usually want to obtain first
results rather quickly. Afterwards, if a final model is estimated the number of iterations of the
sampler is usually set much higher to get close to i.i.d.\ samples from the posterior distribution.
To obtain reasonable starting values for the MCMC sampler we run a backfitting algorithm that
optimizes the posterior mode. Using the main model fitting function \fct{bamlss} all model fitting
engines can be exchanged, which is explained in detail in Section~\ref{sec:package} and the
application Section~\ref{sec:application}. The default model fitting engines use family objects
(see also Section~\ref{sec:package}), similar to the families that can be used
with the \fct{glm} function, which enables easy implementation of new distributions (models).

Note, to capture nonlinearities, a quadratic term for variable \code{age} is added to
the model. The resulting object \code{b} is of class \code{"bamlss"} for which standard
extractor functions like \fct{summary}, \fct{coef}, \fct{plot}, \fct{predict}, etc.\
are available. \hypertarget{msummary}{}The model summary output is printed
by~\label{logitmodel_summary}
\begin{Schunk}
\begin{Sinput}
R> summary(b)
\end{Sinput}
\begin{Soutput}
Call:
bamlss(formula = f, family = "binomial", data = SwissLabor)
---
Family: binomial 
Link function: pi = logit
*---
Formula pi:
---
participation ~ income + age + education + youngkids + oldkids + 
    foreign + I(age^2)
-
Parametric coefficients:
                Mean     2.5%      50%    97.5% parameters
(Intercept)  6.15503  1.55586  5.99204 11.11051      6.196
income      -1.10565 -1.56986 -1.10784 -0.68652     -1.104
age          3.45703  2.05897  3.44567  4.79139      3.437
education    0.03354 -0.02175  0.03284  0.09223      0.033
youngkids   -1.17906 -1.51099 -1.17683 -0.83047     -1.186
oldkids     -0.24122 -0.41231 -0.24099 -0.08054     -0.241
foreignyes   1.16749  0.76276  1.17035  1.55624      1.168
I(age^2)    -0.48990 -0.65660 -0.49205 -0.31968     -0.488
alpha        0.87585  0.32301  0.99408  1.00000         NA
---
Sampler summary:
-
DIC = 1033.325 logLik = -512.7258 pd = 7.8734
runtime = 1.417
---
Optimizer summary:
-
AICc = 1033.737 converged = 1 edf = 8
logLik = -508.7851 logPost = -571.3986 nobs = 872
runtime = 0.012
---
\end{Soutput}
\end{Schunk}
and is based on MCMC samples, which suggest ``significant'' effects for all covariates, 
except for variable \code{education}, since the 95\% credible interval contains zero.
In addition, the acceptance probabilities \code{alpha} are reported and indicate
proper behavior of the MCMC algorithm.
The column \code{parameters} shows respective posterior mode estimates of the regression
coefficients, which are calculated by the upstream backfitting algorithm. Before 
proceeding the analysis, users usually perform additional convergence checks of the
MCMC chains by looking at traceplots and auto-correlation.
\begin{Schunk}
\begin{Sinput}
R> plot(b, which = c("samples", "max-acf"))
\end{Sinput}
\end{Schunk}
\begin{figure}[!ht]
\centering
\setkeys{Gin}{width=1\textwidth}
\includegraphics{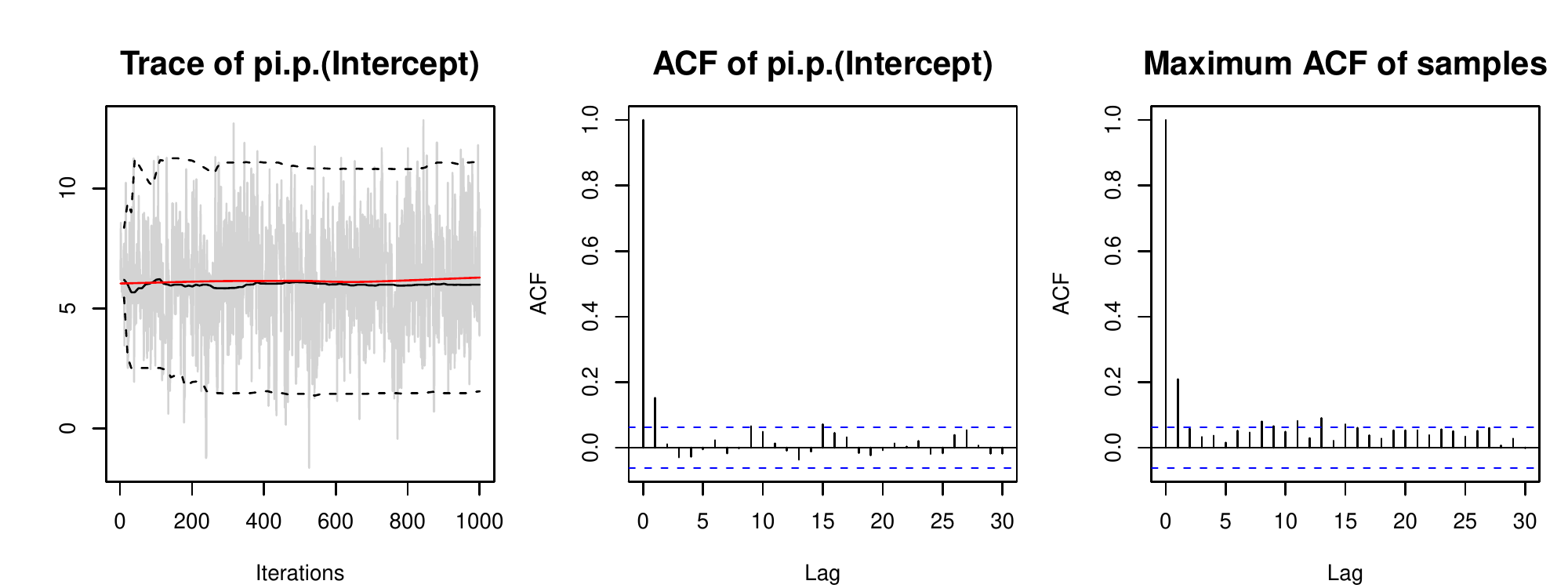}
\caption{\label{fig:logit_traceplots} Logit model, MCMC trace (left panel),
  auto-correlation for the intercept (middle panel), maximum auto-correlation for all
  parameters (right panel).}
\end{figure}
These are visualized in Figure~\ref{fig:logit_traceplots} and reveal approximate convergence of
the MCMC chains, i.e., there is no visible trend and the very low
auto-correlation shown for the intercept and the maximum auto-correlation of all
parameters suggest close to i.i.d.\ samples from the posterior distribution. As mentioned above, the
user could also increase the number iterations and the burnin-phase, as well as adapt the thinning
parameter, to make the significant bar at lag one disappear.
Note that the function call would compute all trace- and auto-correlation plots, however,
for convenience we only show plots for the intercept. In addition, samples can also be
extracted using function \fct{samples}, which returns an object of class \code{"mcmc"},
a class provided by the \pkg{coda} package \citep{bamlss:Plummer+Best+Cowles+Vines:2006}.
This package includes a rich infrastructure for further convergence diagnostic checks,
e.g., Gelman and Rubin's convergence diagnostic
\citep{bamlss:Gelman+Rubin:1992, bamlss:Brooks+Gelman:1998} or
Heidelberger and Welch's convergence diagnostic
\citep{bamlss:Heidelberger+Welch:1981, bamlss:Heidelberger+Welch:1983}.

\begin{figure}[!t]
\centering
\setkeys{Gin}{width=0.4\textwidth}
\includegraphics{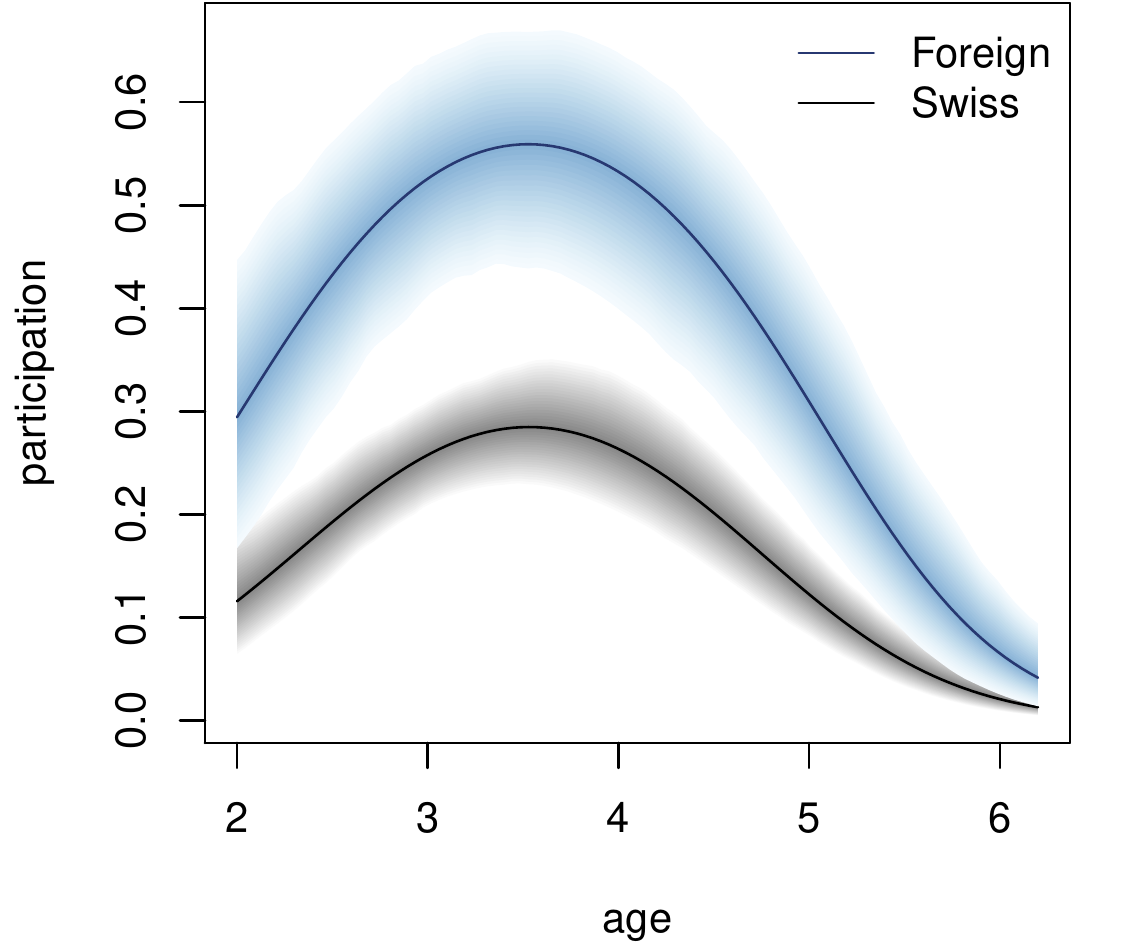}
\includegraphics{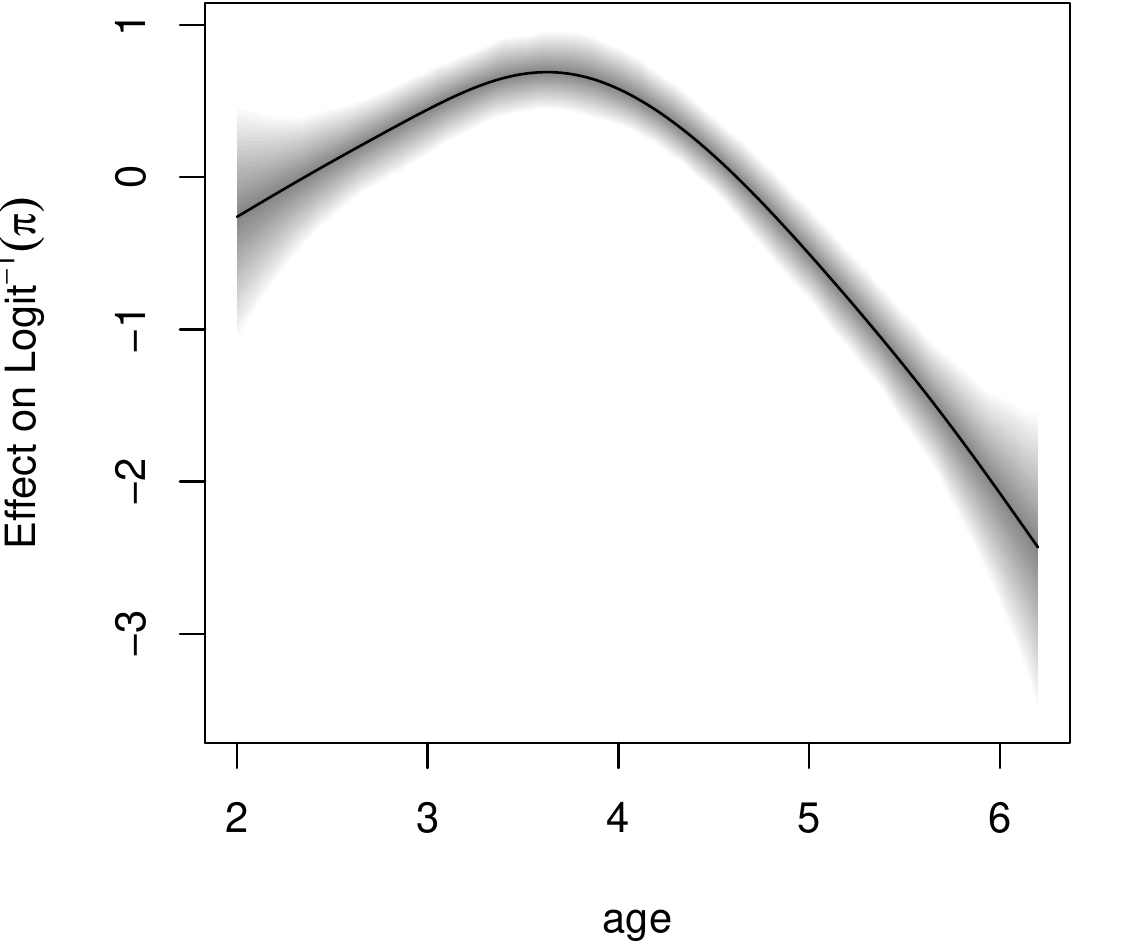}
\caption{\label{fig:logit_effects} Left panel, quadratic polynomial effect of covariate \code{age}
  on estimated probabilities for both cases, \code{foreign} \code{"yes"} and \code{"no"}.
  Right panel, effect on $\text{Logit}^{-1}(\pi)$ of variable \code{age} using regression splines
  (see Section~\ref{sec:flexm}).
  The solid lines represent mean estimates, the shaded areas show 95\% credible intervals.}
\end{figure}

Model predictions on the probability scale can be obtained by the predict method, e.g.,
to visualize the effect of covariate \code{age} on the probability we can create a new
data frame for prediction
\begin{Schunk}
\begin{Sinput}
R> nd <- data.frame(income = 11, age = seq(2, 6.2, length = 100),
+    education = 12, youngkids = 1, oldkids = 1, foreign = "no")
\end{Sinput}
\end{Schunk}
Afterwards, we predict for both cases of variable \code{foreign}
\begin{Schunk}
\begin{Sinput}
R> nd$pSwiss <- predict(b, newdata = nd, type = "parameter", FUN = c95)
R> nd$foreign <- "yes"
R> nd$pForeign <- predict(b, newdata = nd, type = "parameter", FUN = c95)
\end{Sinput}
\end{Schunk}
The predict method is applied on all MCMC samples and argument \code{FUN} specifies a function
that can be applied on the predictor or distribution parameter samples. The default is the \fct{mean} function,
however, in this case we additionally extract the empirical 2.5\% and 97.5\% quantiles
using function \fct{c95} to obtain credible intervals (note, individual samples can be
extracted by passing \code{FUN = identity}, i.e., this way users can
easily generate their own statistics). Then, the estimated effect can be visualized with
\begin{Schunk}
\begin{Sinput}
R> blues <- function(n, ...) sequential_hcl(n, "Blues", rev = TRUE)
R> plot2d(pSwiss ~ age, data = nd, ylab = "participation",
+    ylim = range(c(nd$pSwiss, nd$pForeign)),
+    fill.select = c(0, 1, 0, 1))
R> plot2d(pForeign ~ age, data = nd, add = TRUE,
+    fill.select = c(0, 1, 0, 1), axes = FALSE,
+    s2.col = blues, col.lines = blues(1))
\end{Sinput}
\end{Schunk}
The estimates are shown in Figure~\ref{fig:logit_effects} and suggest a clear difference 
for the effect of \code{age} between both cases of factor variable \code{foreign}.

\subsection{Flexible model terms and estimators} \label{sec:flexm}

Using the flexible infrastructure of \pkg{bamlss}, model terms can be easily exchanged. To give a
first impression of the modeling capabilities, we again use the \code{SwissLabor} data and binomial
logit model of Section~\ref{sec:logitmodel}, however, in this example we use
regression splines to capture the nonlinear effect variable \code{age}.

As noted in the introduction, the \pkg{bamlss} package heavily builds upon the \proglang{R}
package \pkg{mgcv} \citep{bamlss:Wood:2019} infrastructures. To estimate a spline model instead of
a polynomial model for variable \code{age} the model formula only needs to be slightly adapted
\begin{Schunk}
\begin{Sinput}
R> f <- participation ~ income + education +
+    youngkids + oldkids + foreign + s(age, k = 10)
\end{Sinput}
\end{Schunk}
The function \fct{s} is the smooth term constructor from the \pkg{mgcv} package, the default of
\fct{s} are thin-plate regression splines with \code{k = 10} basis functions. The model is again
fitted by
\begin{Schunk}
\begin{Sinput}
R> set.seed(123)
R> b <- bamlss(f, family = "binomial", data = SwissLabor)
\end{Sinput}
\end{Schunk}
The estimated nonlinear effect can be plotted instantly by typing
\begin{Schunk}
\begin{Sinput}
R> plot(b, term = "s(age)")
\end{Sinput}
\end{Schunk}
The estimated effect based on regression splines is shown in the right panel of
Figure~\ref{fig:logit_effects} and reveals that the quadratic polynomial seems to capture the
nonlinearity appropriately. 

To give a better impression what type of model terms can be used with the \pkg{bamlss} framework
Table~\ref{tab:modelterms} lists commonly used specifications.
\begin{table}[t!]
\centering
\begin{tabular}{p{6cm}p{8.3cm}}
\hline
Description & Formula \\ \hline
Linear effects: $\mathbf{X}\boldsymbol{\beta}$ & \code{x1 + x2 + x3} \\ \hline
Nonlinear effects of continuous\newline covariates: $f(\mathbf{x}) = f(x_1)$ & \multirow{2}{*}{\code{s(x1)}} \\ \hline
Two-dimensional surfaces:\newline $f(\mathbf{x}) = f(x_1, x_2)$ & \code{s(x1,x2)}, \code{te(x1,x2)} or \code{ti(x1,x2)} 
\newline(higher dimensional terms possible). \\ \hline
Spatially correlated effects:\newline $f(\mathbf{x}) = f_{spat}(x_s)$ & \code{s(xs, bs = "mrf", xt = list(penalty = K))}, where \code{xs} is a factor indicating the discrete regional information and \code{K} is a supplied penalty matrix. Other options within the \code{xt} argument are possible, please see the documentation of \code{smooth.construct.mrf.smooth.spec()}. \\ \hline
Varying coefficients:\newline $f(\mathbf{x}) = x_1f(x_2)$ & \multirow{2}{*}{\code{s(x2, by = x1)}} \\ \hline
Spatially varying effects:\newline $f(\mathbf{x}) = x_1f_{spat}(x_s)$ or\newline $f(\mathbf{x}) = x_1f(x_2, x_3)$ &
  \code{s(xs, bs = "mrf", xt = list(penalty = K), by = x1)}, \code{s(x2, x3, by = x1)} or\newline \code{te(x2, x3, by = x1)} \\ \hline
Random intercepts with cluster\newline index $c$: $f(\mathbf{x}) = \beta_c$ & \code{s(id, bs = "re")}, where \code{id} is a factor of cluster indices. \\ \hline
Random slopes with cluster index $c$: $f(\mathbf{x}) = x_1\beta_c$ & \code{s(id, x1, bs = "re")}, as above with continuous covariate \code{x1}. \\ \hline
\end{tabular}
\caption{\label{tab:modelterms} Commonly used model term specifications with respective \proglang{R}
  formula syntax.}
\end{table}

Besides the supported infrastructures from the \pkg{mgcv} package, it is also possible to
implement completely new model terms that may follow different setups compared to the basis
functions approach (see also Appendix~\ref{appendix:specialterms} for an example using growth curves).
Moreover, using \pkg{bamlss}, estimation engines can also be exchanged. To give
an example we estimate the nonlinear \code{age} effect in the \code{SwissLabor} example using
a fused lasso algorithm (see also Section~\ref{sec:application} for a complex example using gradient
boosting optimization).
The algorithm performs variable selection in combination with factor fusion (clustering) and can also
be used to identify interpretable nonlinearities. Methodological details on lasso-type penalization
using \pkg{bamlss} are provided in \citet{bamlss:Groll+Hambuckers+Kneib+Umlauf:2019}.
To apply the fused lasso, the numeric variable \code{age} is categorized using empirical quantiles,
e.g., with
\begin{Schunk}
\begin{Sinput}
R> SwissLabor$cage <- cut(SwissLabor$age,
+    breaks = quantile(SwissLabor$age, prob = seq(0, 1, length = 10)),
+    include.lowest = TRUE, ordered_result = TRUE)
\end{Sinput}
\end{Schunk}
The formula for the fused lasso model is then specified with the
special \fct{la} model term constructor function provided in \pkg{bamlss}:

\begin{Schunk}
\begin{Sinput}
R> f <- participation ~ income + education + youngkids + oldkids + foreign + 
+    la(cage, fuse = 2)
\end{Sinput}
\end{Schunk}
where argument \code{fuse} specifies the type of fusion (nominal fusion \code{fuse = 1},
ordered fusion \code{fuse = 2}). To estimate the fused lasso model only the default \code{optimizer}
function in the \fct{bamlss} wrapper function call needs to exchanged
\begin{Schunk}
\begin{Sinput}
R> b <- bamlss(f, family = "binomial", data = SwissLabor,
+    optimizer = lasso, sampler = FALSE,
+    criterion = "BIC", upper = exp(5), lower = 1)
\end{Sinput}
\end{Schunk}
The optimum shrinkage parameter $\lambda$ is selected by the BIC. Arguments \code{upper} and
\code{lower} determine the search interval of $\lambda$, per default \code{nlambda = 100} values
are generated.
Note that no MCMC sampling is used after the \fct{lasso} estimation engine is applied,
argument \code{sampler = FALSE} in the \fct{bamlss} call.

\begin{figure}[t!]
\centering
\setkeys{Gin}{width=0.24\textwidth}
\includegraphics{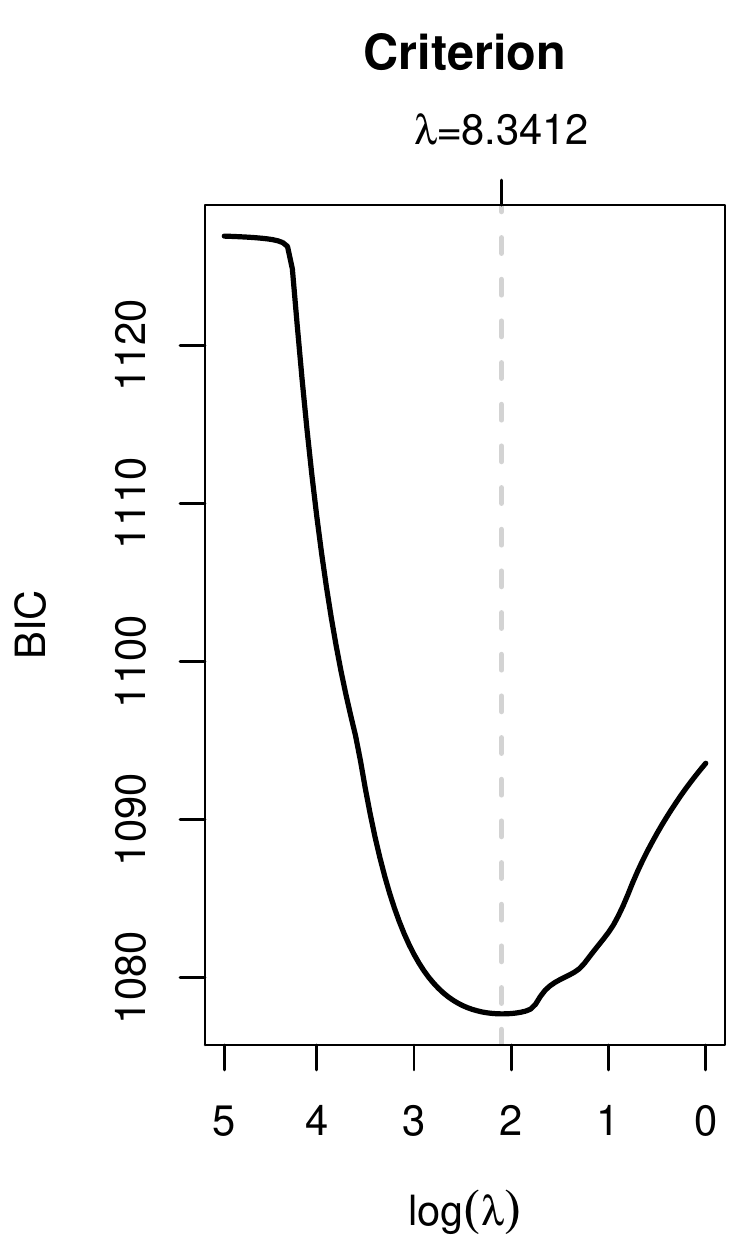}
\setkeys{Gin}{width=0.36\textwidth}
\includegraphics{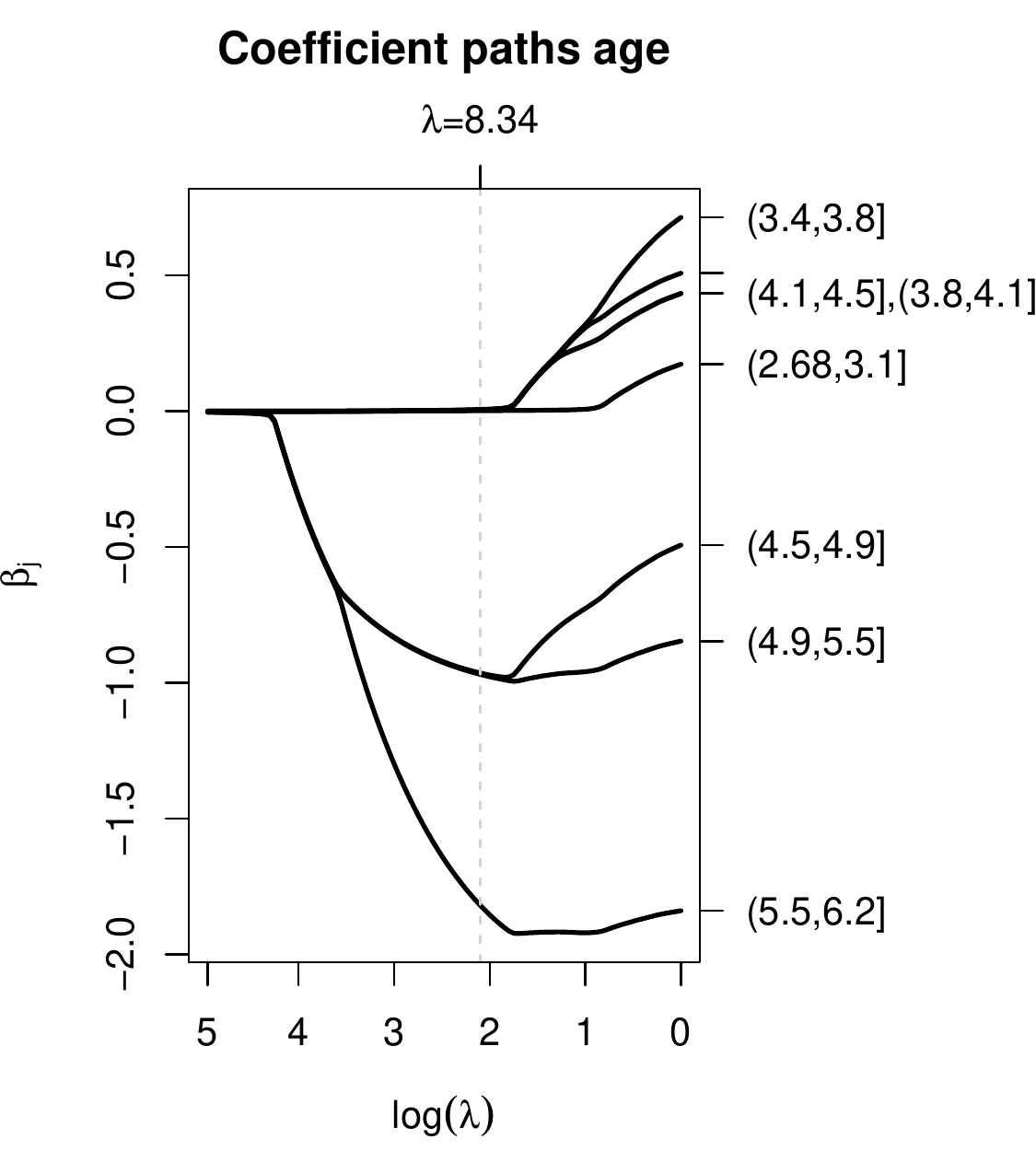}
\setkeys{Gin}{width=0.32\textwidth}
\includegraphics{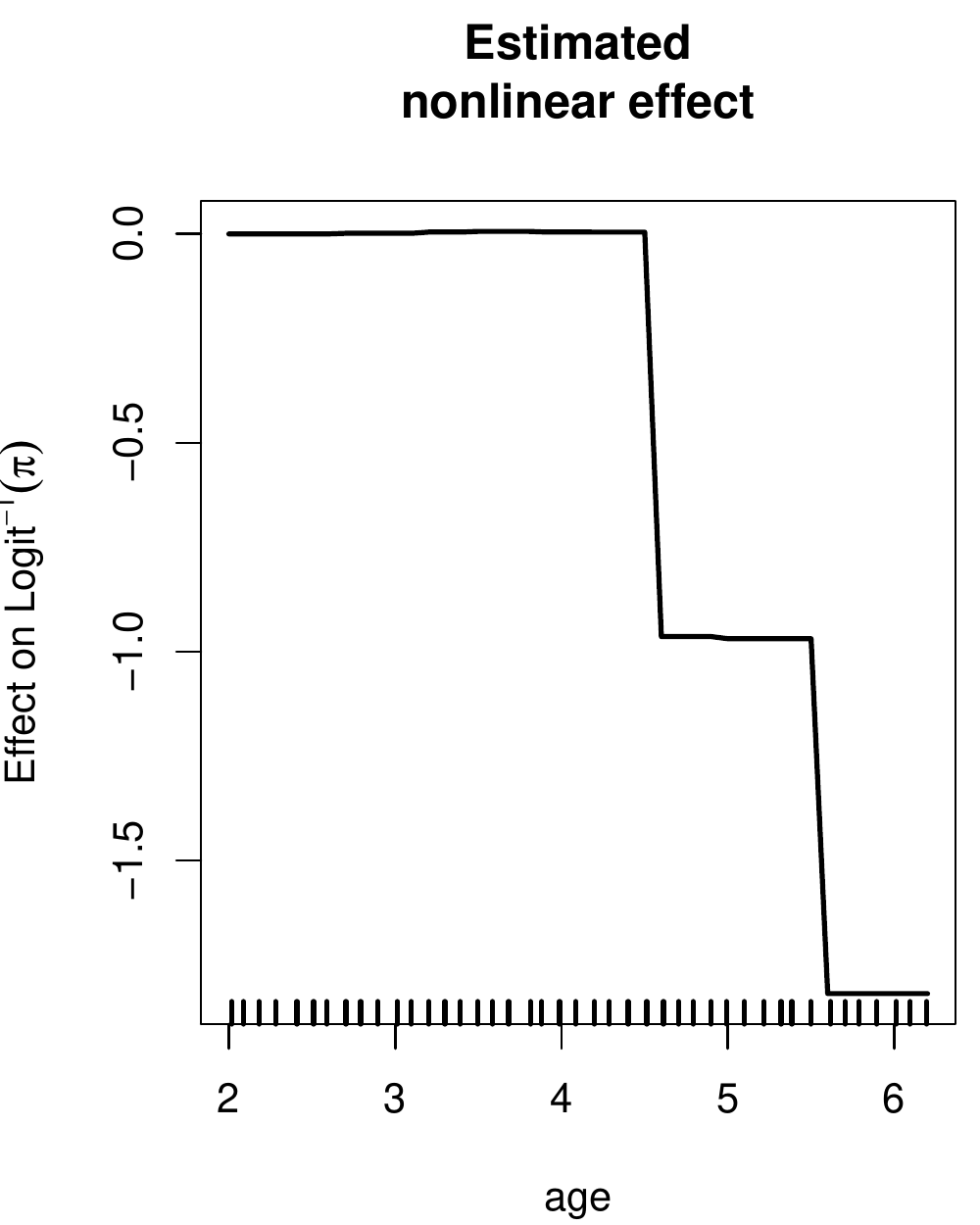}
\caption{\label{fig:lasso} Left panel, BIC curve with optimum shrinkage parameter $\lambda$
  of the lasso example model. The middle panel shows the corresponding coefficient paths for
  variable \code{cage}. The right panel displays the respective estimated effect.}
\end{figure}

The BIC curve and the coefficient paths including the optimum shrinkage parameter $\lambda$ can
be visualized with
\begin{Schunk}
\begin{Sinput}
R> pathplot(b)
\end{Sinput}
\end{Schunk}
Figure~\ref{fig:lasso} shows the BIC curve and coefficient paths for \code{cage}. The BIC curve assumes a clear minimum at the vertical
gray dashed line. The coefficient paths obviously depict that the algorithm can either shrink
categories out of the model (shrink to zero), or even fuses them. In the right panel of
Figure~\ref{fig:lasso}, the estimated effect of the categorized variable
\code{age} is shown. The effect is computed by predicting without
intercept using the optimum stopping iteration, which is selected by BIC and can be extracted with
function \fct{lasso\_stop}. The stopping iteration is passed to the \fct{predict} method by specifying
the \code{mstop} argument.
\begin{Schunk}
\begin{Sinput}
R> page <- predict(b, term = "cage", intercept = FALSE,
+    mstop = lasso_stop(b))
\end{Sinput}
\end{Schunk}
The figure is then created using the untransformed original covariate on the x-axis.
\begin{Schunk}
\begin{Sinput}
R> plot2d(page ~ age, data = SwissLabor, rug = TRUE)
\end{Sinput}
\end{Schunk}
Using the fused lasso estimation some nonlinearities can be identified again, however, the
BIC criterion seems to shrink out the positive effects that are shown for the spline estimate
in the right panel of Figure~\ref{fig:logit_effects}.

\subsection{Location-scale model} \label{sec:locscalemodel}

In this example we will now extend the framework and estimate a complete distributional regression
model using a small textbook example of the well-known simulated motorcycle accident data
\citep{bamlss:Silverman:1985}.
\begin{Schunk}
\begin{Sinput}
R> data("mcycle", package = "MASS")
\end{Sinput}
\end{Schunk}
The data set contains measurements of the head acceleration
(in $g$, variable \code{accel}) in a simulated motorcycle accident, recorded in milliseconds after
impact (variable \code{times}). To estimate a location-scale model with
$$
\texttt{accel} \sim \mathcal{N}(\mu = f_{\mu}(\texttt{times}),
  \log(\sigma) = f_{\sigma}(\texttt{times}))
$$
where functions $f_{\mu}( \cdot )$ and $f_{\sigma}( \cdot )$ are unspecified smooth
functions, which are estimated using regression splines. The log-link for parameter
$\sigma$ ensures positivity. We can use the following model formula for estimation
\begin{Schunk}
\begin{Sinput}
R> f <- list(accel ~ s(times, k = 20), sigma ~ s(times, k = 20))
\end{Sinput}
\end{Schunk}
again, function \fct{s} is the smooth term constructor from the \pkg{mgcv} package
\citep{bamlss:Wood:2019}.
Note that model formulae are provided as lists of formulae, i.e., each list entry
represents one parameter of the response distribution. Moreover, note that all smooth
terms, i.e., \fct{te}, \fct{ti}, etc., are supported by \pkg{bamlss}. This way, it is
also possible to incorporate user defined model terms. A full Bayesian semi-parametric 
distributional regression model can be estimated with
\begin{Schunk}
\begin{Sinput}
R> set.seed(456)
R> b <- bamlss(f, family = "gaussian", data = mcycle)
\end{Sinput}
\end{Schunk}
After the estimation algorithms are finished, the estimated effects can be visualized instantly
using the plotting method.
\begin{Schunk}
\begin{Sinput}
R> plot(b, model = c("mu", "sigma"))
\end{Sinput}
\end{Schunk}
\begin{figure}[!ht]
\centering
\setkeys{Gin}{width=0.4\textwidth}
\includegraphics{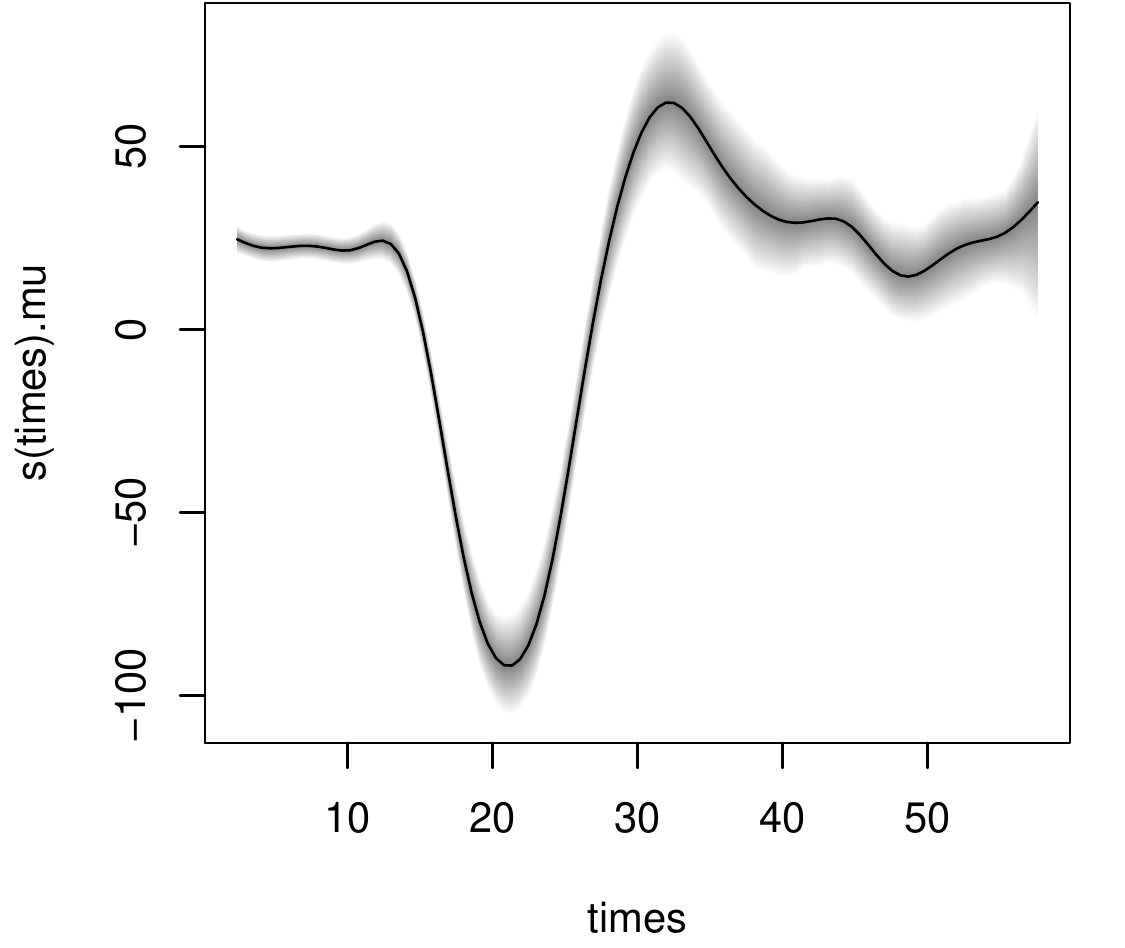}
\includegraphics{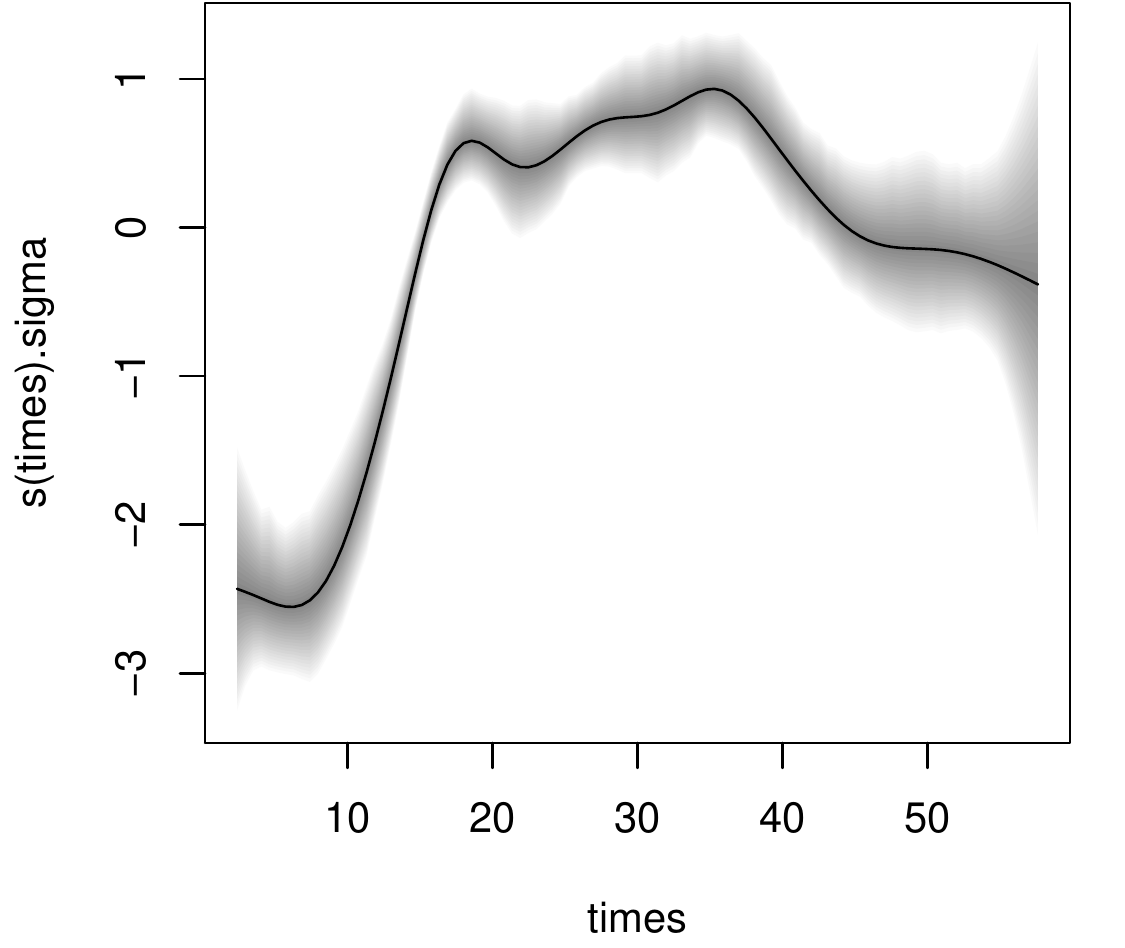}
\caption{\label{fig:toy_effects} Estimated effects of \code{times} on parameter
  $\mu$ and $\sigma$ of the normal location-scale model. The grey shaded areas represent
  95\% credible intervals.}
\end{figure}
The estimated effects are shown in Figure~\ref{fig:toy_effects} depicting a clear nonlinear
relationship for parameter $\mu$ and $\sigma$.

For judging how well the model fits to the data the user can inspect randomized quantile
residuals \citep{bamlss:Dunn+Smyth:1996} using histograms or quantile-quantile plots. 
Residuals can be extracted using function \fct{residuals} and has a plotting method. 
Alternatively, residuals can be investigated with
\begin{Schunk}
\begin{Sinput}
R> plot(b, which = c("hist-resid", "qq-resid"))
\end{Sinput}
\end{Schunk}
\begin{figure}[!ht]
\centering
\setkeys{Gin}{width=0.4\textwidth}
\includegraphics{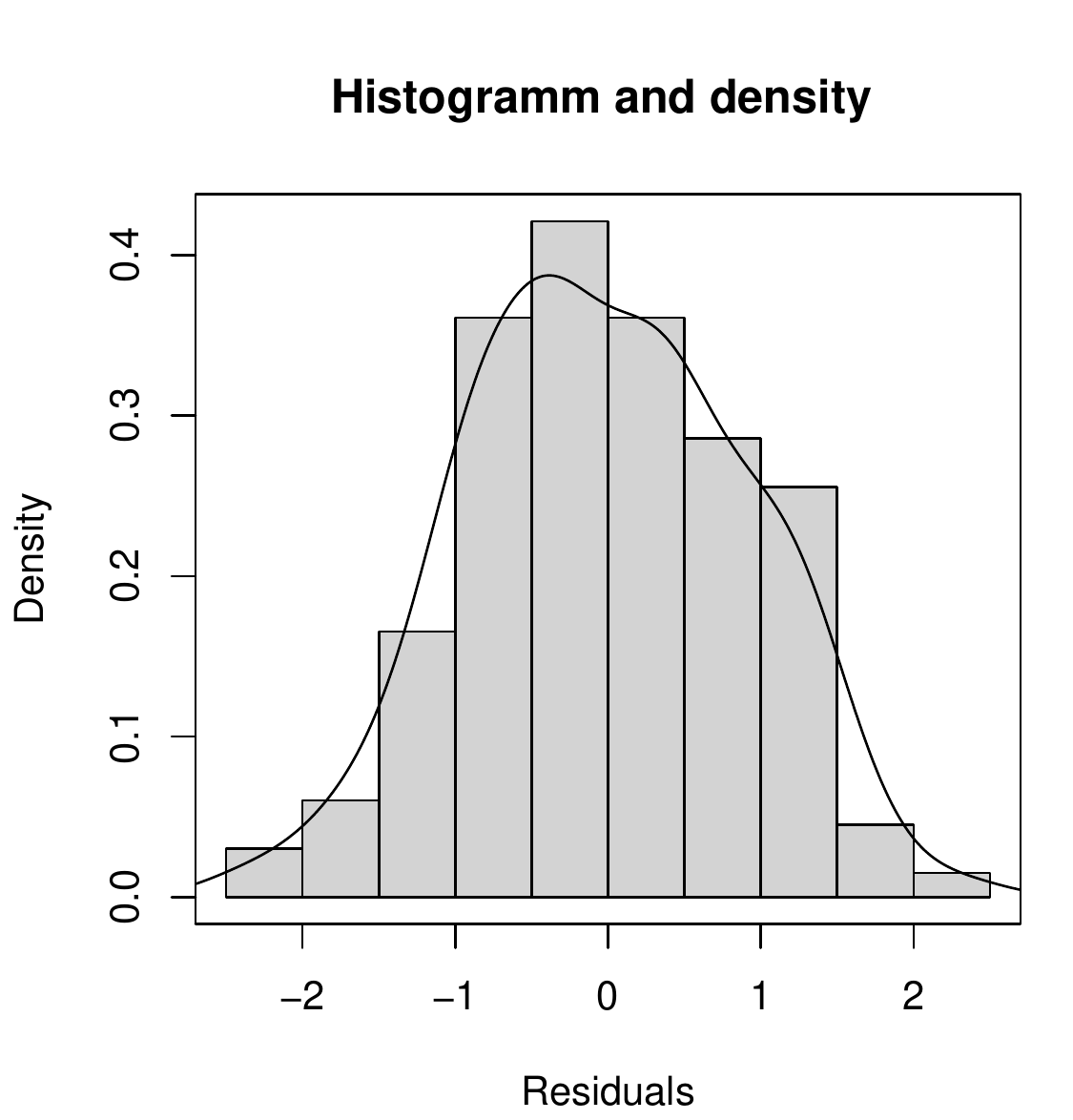}
\includegraphics{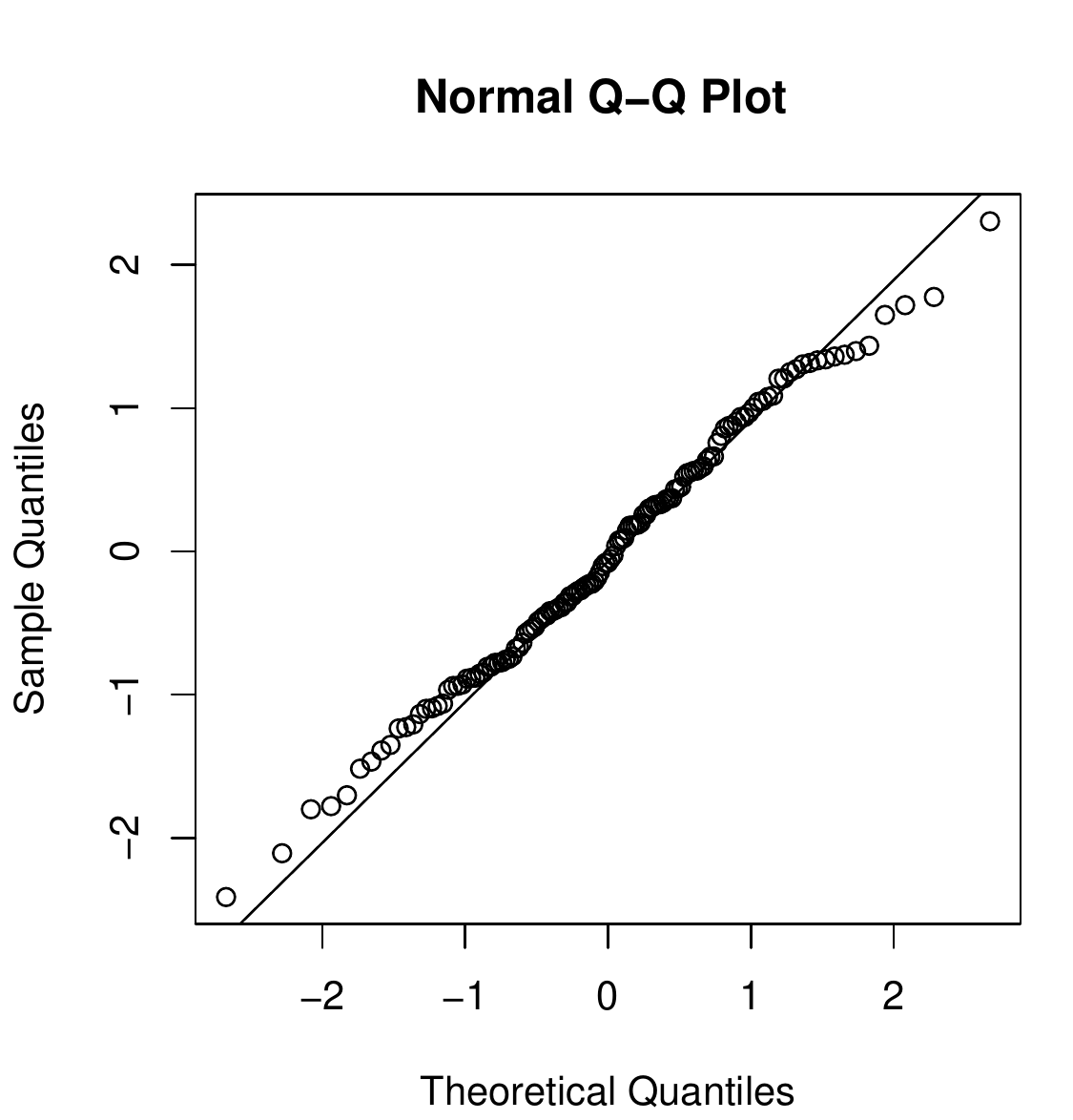}
\caption{\label{fig:toy_resids} Histogram and quantile-quantile plot of the
  resulting randomized quantile residuals of the normal location-scale model.}
\end{figure}
According the histogram and the quantile-quantile plot of the resulting randomized
quantile residuals in Figure~\ref{fig:toy_resids}, the model seems to fit relatively well. 
Only for very low and very high values of \code{accel} the fitted distributions seem
to be less appropriate.

Besides residuals, users can evaluate the model performance, e.g., for model selection 
based on the deviance information criterion (DIC), which can be extracted using function 
\fct{DIC}
\begin{Schunk}
\begin{Sinput}
R> DIC(b)
\end{Sinput}
\begin{Soutput}
      DIC       pd
 1115.247 24.07131
\end{Soutput}
\end{Schunk}
and is also reported in the model summary output. Furthermore, statistical calibration
of fitted models can be assessed by scoring rules
\citep{bamlss:Gneiting+Raftery:2007, bamlss:Gneiting+Balabdaoui+Raftery:2007}. For example, 
the \proglang{R} package \pkg{scoringRules} \citep{bamlss:Jordan+Krueger+Lerch:2018} 
provides easy evaluation of the continuous rank probability score (CRPS) for a couple of 
distributions. Moreover, the Appendix~\ref{appendix:crps} provides a code snippet that computes
the CRPS using numerical integration.

\section{A flexible Bayesian model framework} \label{sec:flexreg}

This section briefly summarizes the BAMLSS modeling framework. For a detailed
methodological description please refer to \citet{bamlss:Umlauf+Klein+Zeileis:2018},
as well as to the \hyperlink{somerefs}{references} given below on page \pageref{somerefs2}
that discuss various applications and extensions that are also implemented in \pkg{bamlss}.
The following outlines the framework from the viewpoint of distributional regression models,
however, please note that model classes like, e.g., GLMs and GAMs or even survival joint models
\citep{bamlss:Koehler+Umlauf+Greven:2016, bamlss:Koehler+Umlauf+Greven:2018}
are special cases in this setup.

\subsection{Model structure}

Within the framework of GAMLSS or distributional regression models
all parameters of the response distribution can be modeled by
explanatory variables such that
\begin{equation} \label{eqn:dreg}
y \sim \mathbf{\mathcal{D}}\left(h_{1}(\theta_{1}) = \eta_{1}, \,\,
  h_{2}(\theta_{2}) = \eta_{2}, \dots, \,\, h_{K}(\theta_{K}) =
  \eta_{K}\right),
\end{equation}
where $\mathbf{\mathcal{D}}$ denotes a parametric distribution for the response
variable $y$ with $K$ parameters $\theta_k$, $k = 1, \ldots, K$, that are linked to 
additive predictors using known monotonic and twice differentiable functions
$h_{k}(\cdot)$. Note that the response may also be a
$q$-dimensional vector $\mathbf{y} = (y_{1}, \ldots, y_{q})^\top$, e.g., when
$\mathbf{\mathcal{D}}$ is a multivariate distribution
(see, e.g., \citealp{bamlss:Klein+Kneib+Klasen+Lang:2015}).
The additive predictor for the $k$-th parameter is given by
\begin{equation} \label{eqn:addpred}
\boldsymbol{\eta}_k = \eta_k(\mathbf{X}; \boldsymbol{\beta}_k) =
  f_{1k}(\mathbf{X}; \boldsymbol{\beta}_{1k}) + \ldots + f_{J_kk}(\mathbf{X}; \boldsymbol{\beta}_{J_kk}),
\end{equation}
based on $j = 1, \ldots, J_k$ unspecified (possibly nonlinear) functions $f_{jk}(\cdot)$, 
applied to each row of the generic data matrix $\mathbf{X}$, encompassing all available 
covariate information. The corresponding parameters
$\boldsymbol{\beta}_k = (\boldsymbol{\beta}_{1k}, \ldots, \boldsymbol{\beta}_{J_kk})^\top$ 
are typically regression coefficients pertaining to model matrices
$\mathbf{X}_k = (\mathbf{X}_{1k}, \ldots, \mathbf{X}_{J_kk})^\top$,
whose structure only depend on the type of covariate(s) and prior assumptions about
$f_{jk}( \cdot )$.

Usually, functions $f_{jk}( \cdot )$ are based on a basis function approach, where $\eta_k$ then
is a typical GAM-type or so-called structured additive predictor
(STAR,~\citealp{bamlss:Fahrmeir+Kneib+Lang:2004}). \citet{bamlss:Umlauf+Klein+Zeileis:2018}
relax this assumption and let $f_{jk}(\cdot)$ be an unspecified composition of covariate data
and regression coefficients. For example, functions $f_{jk}(\cdot)$ could also represent nonlinear
growth curves, a regression tree, a neural network or lasso-penalized model terms as shown
in Section~\ref{sec:flexm}.

For full Bayesian inference, priors need to be assigned to the regression coefficients
$\boldsymbol{\beta}_{jk}$. To be as flexible as possible, \citet{bamlss:Umlauf+Klein+Zeileis:2018}
use the rather general prior $
p_{jk}(\boldsymbol{\beta}_{jk}; \boldsymbol{\tau}_{jk}, \boldsymbol{\alpha}_{jk}) 
$ for the $j$-th model term of the $k$-th parameter, where the form of $p_{jk}( \cdot )$ depends
on the type of function $f_{jk}( \cdot )$. Here,
$\boldsymbol{\tau} =
(\boldsymbol{\tau}_{11}^\top, \ldots, \boldsymbol{\tau}_{J_11}^\top, \ldots,
\boldsymbol{\tau}_{1K}^\top, \ldots, \boldsymbol{\tau}_{J_KK}^\top)^\top$ is
the vector of all assigned hyper-parameters, e.g., representing smoothing variances (shrinkage parameters).
Similarly, $\boldsymbol{\alpha}_{jk}$ is the set of all prior specifications. In most
situations the prior $p_{jk}(\boldsymbol{\beta}_{jk}; \boldsymbol{\tau}_{jk}, \boldsymbol{\alpha}_{jk})$
is based on a multivariate normal kernel for $\boldsymbol{\beta}_{jk}$ and on inverse gamma
distributions for each $\boldsymbol{\tau}_{jk} = (\tau_{1jk}, \ldots, \tau_{L_{jk}jk})^\top$,
but as indicated previously, in principle any type of prior can be used
(see \citealp{bamlss:Gelman:2006, bamlss:Polson+Scott:2012, bamlss:Klein+Kneib:2016,
bamlss:Umlauf+Klein+Zeileis:2018} for
more detailed discussions on priors for $\boldsymbol{\beta}_{jk}$ and $\boldsymbol{\tau}_{jk}$).

\hypertarget{somerefs}{}Examples of distributional models that fit well in this
framework are the ones for: \label{somerefs2}
\begin{itemize}
  \item Univariate responses of any type, e.g.~counts with zero-inflation and or overdispersion as
    proposed in \citet{bamlss:Klein+Kneib+Lang:2015, bamlss:Herwartz+Klein+Strumann:2016},
    continuous responses with spikes, skewness, heavy tails or bounded support as in
    \citet{bamlss:Klein+Kneib+Lang+Sohn:2015,bamlss:Klein+Denuit+Kneib+Lang:2014}, as well as
    responses for extreme events \citep{bamlss:Umlauf+Kneib:2018}.
 \item Multivariate responses such as multivariate normal, multivariate t or Dirichlet regression
   (for analyzing compositional data, \citealp{bamlss:Klein+Kneib+Klasen+Lang:2015}).
 \item Multivariate responses with more complex dependence structures modeled through copulas
   \citet{bamlss:Klein+Kneib:2016b}.
 \item Survival data and joint modeling~\citep{bamlss:Koehler+Umlauf+Greven:2016, bamlss:Koehler+Umlauf+Greven:2018}.
\end{itemize}

\subsection{Posterior estimation} \label{sec:algos}

Estimation typically requires to evaluate the log-likelihood
$\ell(\boldsymbol{\beta}; \mathbf{y}, \mathbf{X})$ function and its derivatives w.r.t.\
all regression coefficients $\boldsymbol{\beta}$ a number of times. For fully Bayesian inference
the log-posterior is either used for posterior mode estimation, or for solving
high-dimensional integrals. e.g., for posterior mean estimation MCMC samples need to be computed.

Although the types of models that can be fitted within the flexible BAMLSS framework
can be quite complex, \citet{bamlss:Umlauf+Klein+Zeileis:2018} show that there are a number
of similarities between optimization and sampling concepts. Fortunately, and albeit the different
model term complexity, algorithms for posterior mode and
mean estimation can be summarized into a partitioned updating scheme with
separate updating equations using leapfrog or zigzag iteration \citep{bamlss:Smyth:1996}, e.g.,
with updating equations
\begin{equation} \label{eqn:blockblockupdate}
(\boldsymbol{\beta}_{jk}^{(t + 1)}, \boldsymbol{\tau}_{jk}^{(t + 1)}) =
  U_{jk}(\boldsymbol{\beta}_{jk}^{(t)}, \boldsymbol{\tau}_{jk}^{(t)}; \, \cdot \,) \qquad
    j = 1, \ldots, J_k, \quad k = 1, \ldots, K,
\end{equation}
where function $U_{jk}( \cdot )$ is an updating function, e.g., for generating one Newton-Raphson
step or for getting the next step in an MCMC simulation, a.o.

Using a basis function approach, the updating functions $U_{jk}( \cdot )$ for posterior mode
(frequentist penalized likelihood) estimation or MCMC for $\boldsymbol{\beta}_{jk}$ share an iteratively
weighted least squares updating step (IWLS,~\citealp{bamlss:Gamerman:1997})
\begin{equation} \label{eqn:blockbackfit}
  \boldsymbol{\beta}_{jk}^{(t+1)}
  = U_{jk}(\boldsymbol{\beta}_{jk}^{(t)}; \, \cdot \,) =
    (\mathbf{X}_{jk}^\top\mathbf{W}_{kk}\mathbf{X}_{jk} +
      \mathbf{G}_{jk}(\boldsymbol{\tau}_{jk}))^{-1}\mathbf{X}_{jk}^\top\mathbf{W}_{kk}(
      \mathbf{z}_k - \boldsymbol{\eta}_{k, -j}^{(t+1)}),
\end{equation}
with weight matrices $\mathbf{W}_{kk}$ and working responses $\mathbf{z}_k$, similarly to the
well-known IWLS updating scheme for generalized linear models (GLM,~\citealp{bamlss:Nelder+Wedderburn:1972}).
The matrices $\mathbf{G}_{jk}(\boldsymbol{\tau}_{jk})$ are derivative matrices of the priors
$p_{jk}(\boldsymbol{\beta}_{jk}; \boldsymbol{\tau}_{jk}, \boldsymbol{\alpha}_{jk})$ w.r.t.\ the
regression coefficients $\boldsymbol{\beta}_{jk}$, e.g.,
$\mathbf{G}_{jk}(\boldsymbol{\tau}_{jk})$ can be a penalty matrices that penalizes the complexity of
$f_{jk}( \cdot )$ using a P-spline representation \citep{bamlss:Eilers+Marx:1996}.

Even if the functions $f_{jk}( \cdot )$ are not based on a basis function approach, the
updating scheme (\ref{eqn:blockbackfit}) can be further generalized to
$$
\boldsymbol{\beta}_{jk}^{(t + 1)} = U_{jk}\left(\boldsymbol{\beta}_{jk}^{(t)},
  \mathbf{z}_k - \boldsymbol{\eta}_{k, -j}^{(t+1)}; \, \cdot \,\right),
$$
i.e., theoretically any updating function applied to the ``partial residuals''
$\mathbf{z}_k - \boldsymbol{\eta}_{k, -j}^{(t+1)}$ can be used. (For detailed derivations see
\citealp{bamlss:Umlauf+Klein+Zeileis:2018}.)

The great advantage of this modular architecture is, that the concept does not limit to modeling
of the distributional parameters $\theta_k$ in (\ref{eqn:dreg}), e.g.\ as mentioned above,
based on the survival function,
\citet{bamlss:Koehler+Umlauf+Greven:2016} and \citet{bamlss:Koehler+Umlauf+Greven:2018} implement
Bayesian joint models for survival and longitudinal data. Moreover, the updating schemes do not
restrict to any particular estimation engine, e.g.,
\citet{bamlss:Groll+Hambuckers+Kneib+Umlauf:2019} use the framework to implement lasso-type
penalization for GAMLSS and \citet{bamlss:Simon+Fabsic+Mayr+Umlauf+Zeileis:2018}
investigate gradient boosting with stability selection algorithms
(see also Section~\ref{sec:application}). Very recently, \citet{bamlss:Umlauf+Klein:2019}
implement neural network distributional regression models.

\subsection{Model choice and evaluation}

\subsubsection{Measures of performance}

Model choice and variable selection is important in distributional regression due to the large
number of candidate models. The following lists commonly used tools:
\begin{itemize}
\item \emph{Information criteria} can be used to compare different model specifications.  
  For posterior mode estimation, the Akaike information criterion (AIC), or the corrected
  AIC, as well as the Bayesian information criterion (BIC), can be used. Estimation of model
  complexity is based on the so-called equivalent degrees of freedom (EDF). 

  For MCMC based estimation, model choice mainly relies on the deviance information criterion
  (DIC, \citealp{SpiBesCarLin2002}) and the widely applicable information criterion
  (WAIC, \citealp{watanabe2010asymptotic}). 
\item \emph{Quantile residuals} \citep{bamlss:Dunn+Smyth:1996} can be used to evaluate the model
  fit. The residuals can be assessed by quantile-quantile-plots, probability integral transforms
  (PIT) histograms \citep{bamlss:Gneiting+Balabdaoui+Raftery:2007} or worm plots
  \citep{bamlss:Buuren+Miranda:2001}.
\item \emph{Scoring rules}: Sometimes it is helpful to evaluate the performance on a test data set
  (or for instance based on cross validation). For this, proper scoring rules
  \citep{bamlss:Gneiting+Raftery:2007, bamlss:Gneiting+Balabdaoui+Raftery:2007} can be utilized.
\end{itemize}

\subsubsection{Evaluation and interpretation}
 \begin{itemize}
\item \emph{Plotting}: Estimated functions $\hat{f}_{jk}( \cdot )$ are usually centered around their
  mean, therefore, simple effect plots are a straightforward method to evaluate individual model
  term importance and can also be used for respective interpretations. Sometimes it can be useful in
  distributional regression to look at transformations of the original model parameters such as
  expected value or variance of the response variable $\mathbf{y}$.
\item \emph{Predictions}: For obtaining such transformations model predictions need to
  be computed. This can be done either manually by the corresponding \fct{predict} method, or by
  the \proglang{R} package \pkg{distreg.vis} \citep{bamlss:distreg.vis}, which provides a
  graphical user interface for visualization of distributional regression models.
 \end{itemize}

\section[The bamlss package]{The \pkg{bamlss} package} \label{sec:package}

The \proglang{R} package \pkg{bamlss} provides a modular software architecture for
flexible Bayesian regression models (and beyond). The implementation follows the 
conceptional framework presented in \citet{bamlss:Umlauf+Klein+Zeileis:2018}, which 
supports Bayesian and/or frequentist estimation engines using complex possibly nonlinear 
model terms of any type. The highlights of the package are:
\begin{itemize}
\item A unified model description where a \code{formula} specifies how to set up the predictors
  from the \code{data} and the \code{family}, which holds information about the response
  distribution, the model.
\item A generic method for setting up model terms and a \fct{model.frame} for BAMLSS, the
  \fct{bamlss.frame}, along with the corresponding prior structures. A \fct{transform}
  function can
  optionally set up modified terms, e.g., using mixed model representation for smooth terms.
\item Support for modular and exchangeable updating functions or complete model fitting engines
  in order to optionally implement either algorithms for maximization of the log-posterior for
  posterior mode estimation or for solving high-dimensional integrals, e.g., for posterior mean
  or median estimation.
  First, an (optional) \fct{optimizer} function can be run, e.g., for computing posterior mode
  estimates. Second, a \fct{sampler} is employed for full Bayesian inference with MCMC, which uses the
  posterior mode estimates from the \fct{optimizer} as staring values. An additional step can be used
  for preparing the \fct{results}, e.g., for creating model term effect plots.
\item Standard post-modeling extractor functions to create sampling statistics, visualizations,
  predictions, etc.
\end{itemize}
\begin{figure}[p!]
\centering
\begin{tikzpicture}[node distance=1.5cm,
    every node/.style={fill=white, font=\sffamily}, align=center]
  \node (data)        [input,fill=rb1]                               {\texttt{data}};
  \node (formula)     [input, left of=data, xshift=-4.5cm,fill=rb1]  {\texttt{formula}};
  \node (family)      [input, right of=data, xshift=-4.5cm,fill=rb1]   {\texttt{family}};

  \begin{pgfonlayer}{background}
  \node[outer,fit=(data) (formula) (family)] (A) {};
  \node (text) [right of=A, xshift=-6.5cm] {Input};
  \end{pgfonlayer}

  \node (bframe) [processing, below of=family, yshift=-0.3cm,fill=rb2] {\texttt{bamlss.frame()}};
  \node (trans)  [processing, below of=bframe, yshift=0.3cm,fill=rb2] {\texttt{transform()}};

  \coordinate[below of=family, yshift=0.6cm] (invisible6);
  \draw[->, shorten >=3pt] (invisible6) -- (bframe);

  \draw[] (data)    -- (invisible6);
  \draw[] (family)  -- (invisible6);
  \draw[] (formula) -- (invisible6);
  \draw[->, shorten >=1pt, line width=0.8mm, gray] (bframe)  -- (trans);

  \begin{pgfonlayer}{background}
  \node[outer,fit=(bframe) (trans), minimum width=8.55cm] (B) {};
  \node (text) [right of=B, xshift=-7.2cm] {Pre-processing};
  \end{pgfonlayer}

  \node (opt) [estimation, below of=trans,fill=rb3] {\texttt{optimizer()}};
  \node (sampler) [estimation, below of=opt, yshift=0.3cm,fill=rb3] {\texttt{sampler()}};

  \coordinate[right of=sampler, yshift=0.15cm] (invisible5);

  \draw[->, shorten >=3pt, line width=0.8mm, gray] (trans) -- (opt);
  \draw[->, shorten >=1pt, line width=0.8mm, gray] (opt)   -- (sampler);
  \draw[->, shorten >=3pt, line width=0.8mm, gray] (trans.west) to [out=-180,in=180] (sampler.west);
  \draw[->, shorten >=6pt, line width=0.8mm, gray] (bframe.east) to [out=10,in=10] (invisible5);
  \draw[->, shorten >=3pt, line width=0.8mm, gray] (bframe.east) to [out=10,in=10] (opt.east);

  \begin{pgfonlayer}{background}
  \node[outer,fit=(opt) (sampler), minimum width=8.55cm] (C) {};
  \node (text) [right of=C, xshift=-6.9cm] {Estimation};
  \end{pgfonlayer}

  \node (sstats) [stats, below of=sampler,fill=rb4] {\texttt{samplestats()}};
  \node (results) [stats, below of=sstats, yshift=0.3cm,fill=rb4] {\texttt{results()}};

  \begin{pgfonlayer}{background}
  \node[outer,fit=(sstats) (results), minimum width=8.55cm] (D) {};
  \node (text) [right of=D, xshift=-7.25cm] {Post-processing};
  \end{pgfonlayer}

  \draw[->, shorten >=3pt, line width=0.8mm, gray] (sampler) -- (sstats);
  \draw[->, shorten >=1pt, line width=0.8mm, gray] (sstats)  -- (results);

  \node (plot) [output, below of=results, yshift=-0.5cm,fill=rb5] {\texttt{plot()}};
  \node (summary) [output, left of=plot, xshift=-1.5cm,fill=rb5]  {\texttt{summary()}};
  \node (predict) [output, right of=plot, xshift=1.5cm,fill=rb5]  {\texttt{predict()}};

  \begin{pgfonlayer}{background}
  \node[outer,fit=(summary) (plot) (predict)] (E) {};
  \node (text) [right of=E, xshift=-6.6cm] {Output};
  \end{pgfonlayer}

  \coordinate[below of=results, yshift=0.55cm] (invisible);
  \coordinate[below of=results, yshift=0.7cm] (invisible3);

  \draw[->, shorten >=3pt] (invisible) -- (plot);
  \draw[->, shorten >=3pt] (invisible) -- (summary);
  \draw[->, shorten >=3pt] (invisible) -- (predict);
  \draw[] (results) -- (invisible);

  \draw[->, shorten >=3pt, line width=0.8mm, gray] (opt.west) to [out=-180,in=180] (results.west);

  \coordinate[left of=opt, xshift=-2cm] (invisible2);
  \coordinate[right of=sampler, xshift=2cm] (invisible4);
  \draw[line width=0.8mm, gray] (opt.west) |- (invisible2);
  \draw[line width=0.8mm, gray] (sampler.east) |- (invisible4);
  \draw[->, shorten >=3pt, line width=0.8mm, gray] (invisible2) |- (invisible3);
  \draw[->, shorten >=3pt, line width=0.8mm, gray] (invisible4) |- (invisible3);

  \draw[->, shorten >=3pt, line width=0.8mm, gray] (sampler.east) to [out=10,in=10] (results.east);

  \end{tikzpicture}
\caption{\label{fig:flowchart} Flow chart of the \pkg{bamlss} modeling architecture. Thick gray
  lines represent optional paths, e.g., after building the \fct{bamlss.frame} the user can either
  run an \fct{optimizer} function prior running the \fct{sampler}, or run the \fct{sampler}
  function directly.}
\end{figure}
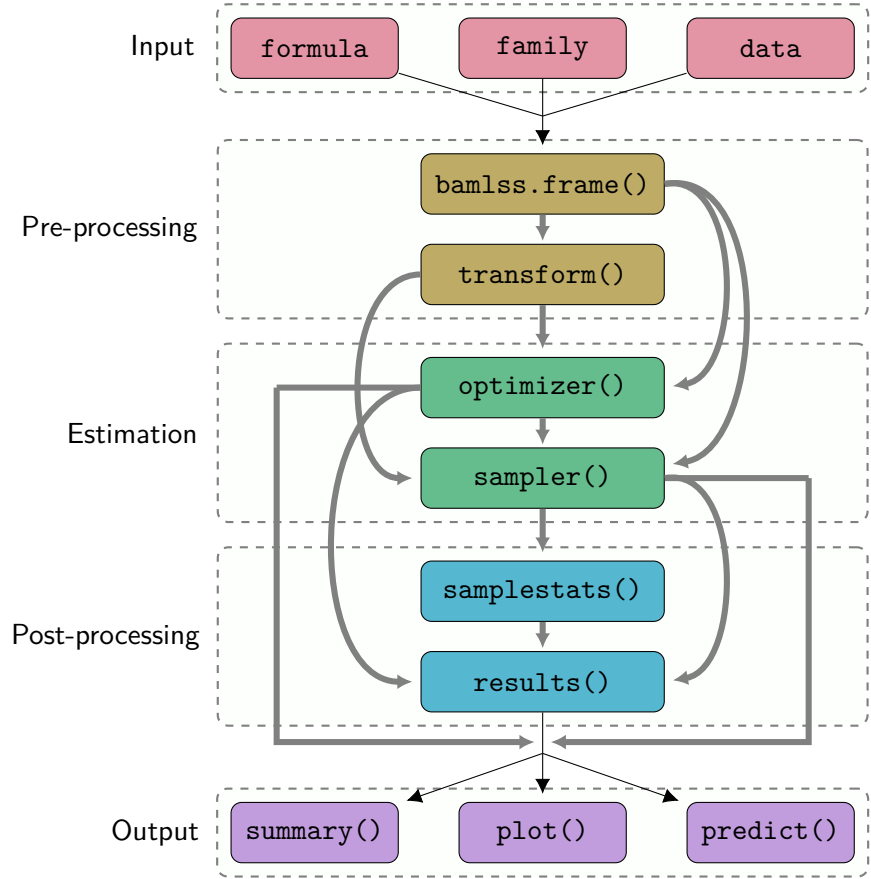
\begin{table}[p!]
\centering
\begin{tabular}{lll}
\hline
Step & Type        & Function \\ \hline
\multirow{3}{*}{Pre-processing} & Parser      & \code{bamlss.frame()} \\ \cline{2-3}
 & \multirow{2}{*}{Transformer} & \code{bamlss.engine.setup()}, \code{randomize()} \\
 & & \code{lasso\_transform()} \\ \hline
\multirow{4}{*}{Estimation} & \multirow{2}{*}{Optimizer} & \code{bfit()}, \code{bbfit()},
                              \code{boost()}, \code{lasso()} \\
 & &\code{cox\_mode()}, \code{jm\_mode()}  \\ \cline{2-3}
 & \multirow{2}{*}{Sampler} & \code{GMCMC()}, \code{BayesX()}, \code{JAGS()} \\
 & & \code{cox\_mcmc()}, \code{jm\_mcmc()} \\ \hline
Post-processing & Stats \& Results & \code{samplestats()}, \code{results.bamlss.default()} \\ \hline
\end{tabular}
\caption{\label{tab:functions} Current available functions that can be used for
  pre-processing, estimation and post-processing within the \pkg{bamlss} framework.}
\end{table}
The modular architecture of \pkg{bamlss} is illustrated in Figure~\ref{fig:flowchart}. As
mentioned above, the first
step in model development is to setup design and penalty matrices for a model that is specified
by the \code{family} object. Therefore a \code{formula} is processed together with the \code{data}
using the \fct{bamlss.frame} function. In a second pre-processing step, the returned model frame
may also be transformed. The BAMLSS model frame can then be used with \fct{optimizer} and/or
\fct{sampler} functions in the estimation step. This is probably the main advantage of the
architecture, users can easily exchange and integrate user defined estimation functions.
The only requirement is to keep the structure of the \fct{bamlss.frame} function, as well for
\fct{optimizer} and \fct{sampler} functions. After the estimation step optional post-processing
functions can be applied to create additional sampling statistics, function \fct{samplestats},
or results that can be used for plotting the estimated effects, function \fct{results}. Note
that the post-processing step is optional since it is not necessarily needed in the last
output step, e.g., for computing predictions. This feature is especially important when using
large data sets, because the run time for computing \fct{samplestats} or \fct{results} can be
quite long or computations can even lead to memory problems. In summary, the architecture is very flexible
such that users interested in implementing new models only need to focus on the
estimation step, i.e., write \fct{optimizer} or \fct{sampler} functions and get all post-processing
and extractor functionalities ``for free''. This way, prototyping
becomes relatively easy, but also the integration/implementation of (new) high-performance
estimation engines is facilitated.
Table~\ref{tab:functions} provides an overview of current available functions.

To exemplify the presented ``Lego toolbox'', the following \proglang{R} code estimates the
logit model using the \code{SwissLabor} data presented in Section~\ref{sec:logitmodel}. First,
the data is loaded and the model formula is specified with
\begin{Schunk}
\begin{Sinput}
R> data("SwissLabor", package = "AER")
R> f <- participation ~ income + age + education +
+    youngkids + oldkids + foreign + I(age^2)
\end{Sinput}
\end{Schunk}
In the second step, the necessary design matrices are constructed using the model frame
parser function \fct{bamlss.frame}
\begin{Schunk}
\begin{Sinput}
R> bf <- bamlss.frame(f, data = SwissLabor, family = "binomial")
\end{Sinput}
\end{Schunk}
Then, posterior mode estimates are obtained by using the implemented backfitting estimation
function \fct{bfit}
\begin{Schunk}
\begin{Sinput}
R> pm <- with(bf, bfit(x, y, family))
\end{Sinput}
\end{Schunk}
The estimated parameters returned from function \fct{bfit} can then be used as starting
values for the MCMC sampler function \fct{GMCMC}
\begin{Schunk}
\begin{Sinput}
R> set.seed(123)
R> samps <- with(bf, GMCMC(x, y, family, start = pm$parameters))
\end{Sinput}
\end{Schunk}
Using the parameters samples returned from function \fct{GMCMC}, statistics like the DIC are
computed using the \fct{samplestats} function
\begin{Schunk}
\begin{Sinput}
R> stats <- with(bf, samplestats(samps, x, y, family))
R> print(unlist(stats))
\end{Sinput}
\begin{Soutput}
    logLik        DIC         pd 
-512.72579 1033.32501    7.87343 
\end{Soutput}
\end{Schunk}
As one can see in the code above, estimation engines have common arguments \code{x} (holding the
design and penalty matrices), \code{y} (the response data) and \code{family} (the \pkg{bamlss}
family object). For implementing new estimation engines, users only need to keep the argument
structures and the return values, i.e., for \fct{optimizer} functions a named numeric vector
of estimated parameters and for \fct{sampler} functions parameter samples of class \code{"mcmc"}
or \code{"mcmc.list"} (see package \pkg{coda}, \citealp{bamlss:Plummer+Best+Cowles+Vines:2006}).
More details on the naming convention and the structure of the return value of \fct{bamlss.frame}
are given in Section~\ref{sec:bf}.

To ease the modeling process, all the single modeling steps presented in the above can be executed
using the \pkg{bamlss} wrapper function \fct{bamlss}. The main arguments of \fct{bamlss} are
\begin{Code}
  bamlss(formula, family = "gaussian", data = NULL,
    transform = NULL,                         ## Pre-processing
    optimizer = NULL, sampler = NULL,         ## Estimation
    samplestats = NULL, results = NULL, ...)  ## Post-processing
\end{Code}
where the first line basically represents the standard model frame specifications
\citep[see][]{bamlss:Chambers+Hastie:1992}. All other arguments represent functions presented
in Table~\ref{tab:functions} and can be exchanged. Note that the default for argument
\code{optimizer} is the backfitting estimation function \fct{bfit} and the default for argument
\code{sampler} is the \fct{GMCMC} sampling function.

The returned fitted model object is a list of class \class{bamlss}, which is supported by several
standard methods and extractor functions, such as \fct{plot}, \fct{summary} and \fct{predict}.

As already exemplified in Section~\ref{sec:motivation}, using the model fitting wrapper function
\fct{bamlss} it is straightforward to use different modeling approaches by simply exchanging
the estimation engines. This feature can be particularly important in complex modeling situation,
where good mixing of the MCMC algorithm requires very good starting values. One use case is presented
in Section~\ref{sec:application}, where for stability reasons posterior mode estimates are obtained
using the gradient boosting optimizer function \fct{boost}. Afterwards the MCMC sampling engine
\fct{GMCMC} is applied with the boosting estimates as starting values.

\subsection{The BAMLSS model frame} \label{sec:bf}

Similar to the well-known \fct{model.frame} function that is used, e.g., by the linear model 
fitting function \fct{lm}, or for generalized linear models \fct{glm}, the \fct{bamlss.frame}
function extracts a ``model frame'' for fitting distributional regression models. 
Internally, the function parses model formulae, one for each parameter of the distribution,
using the \pkg{Formula} package infrastructures \citep{bamlss:Zeileis+Croissant:2010} in combination
with \fct{model.matrix} processing for linear effects and \fct{smooth.construct} processing of
the \pkg{mgcv} package to setup design and penalty matrices for unspecified smooth function 
estimation (\citealp{bamlss:Wood:2019}, see also, e.g., the documentation of function \fct{s} and
\fct{te}).

The most important arguments are
\begin{Code}
  bamlss.frame(formula, data = NULL, family = "gaussian",
    weights = NULL, subset = NULL, offset = NULL,
    na.action = na.omit, contrasts = NULL, ...)
\end{Code}
The argument \code{formula} can be a classical model formulae, e.g., as used by the \fct{lm} function,
or an extended \pkg{bamlss} formula including smooth term specifications like \fct{s} or \fct{te},
that is internally parsed by function \fct{bamlss.formula}.
Note that the \pkg{bamlss} package uses special \code{family} objects, that can be passed either as
a character without the \code{"\_bamlss"} extension of the \pkg{bamlss} family name (see the manual
\code{?bamlss.family} for a list of available families), or the family function itself.
In addition, all families of the \pkg{gamlss} \citep{bamlss:Stasinopoulos+Rigby:2019} and
\pkg{gamlss.dist} \citep{bamlss:gamlss.dist} package are supported.

The returned object, a named list of class \code{"bamlss.frame"}, can be employed with the model
fitting engines listed in Table~\ref{tab:functions}. The most important elements used for
estimation are:
\begin{itemize}
\item \code{x}: A named list, the elements correspond to the parameters that are specified within
  the \code{family} object. For each distribution parameter, the list contains all design and
  penalty matrices needed for modeling (see the upcoming example).
\item \code{y}: The response data.
\item \code{family}: The processed \pkg{bamlss} \code{family}.
\end{itemize}
To better understand the structure of the \code{"bamlss.frame"} object a print method is provided.
For illustration, we simulate data
\begin{Schunk}
\begin{Sinput}
R> set.seed(111)
R> d <- GAMart()
\end{Sinput}
\end{Schunk}
and set up a \code{"bamlss.frame"} object for a Gaussian
distributional regression model including smooth terms. First, a model formula is needed
\begin{Schunk}
\begin{Sinput}
R> f <- list(
+    num ~ x1 + s(x2) + s(x3) + te(lon,lat),
+    sigma ~ x1 + s(x2) + s(x3) + te(lon,lat)
+  )
\end{Sinput}
\end{Schunk}
Afterwards the model frame can be computed with
\begin{Schunk}
\begin{Sinput}
R> bf <- bamlss.frame(f, data = d, family = "gaussian")
\end{Sinput}
\end{Schunk}
To keep the overview, there is also an implemented print method for \code{"bamlss.frame"} objects.
\begin{Schunk}
\begin{Sinput}
R> print(bf)
\end{Sinput}
\begin{Soutput}
'bamlss.frame' structure: 
  ..$ call 
  ..$ model.frame 
  ..$ formula 
  ..$ family 
  ..$ terms 
  ..$ x 
  .. ..$ mu 
  .. .. ..$ formula 
  .. .. ..$ fake.formula 
  .. .. ..$ terms 
  .. .. ..$ model.matrix 
  .. .. ..$ smooth.construct 
  .. ..$ sigma 
  .. .. ..$ formula 
  .. .. ..$ fake.formula 
  .. .. ..$ terms 
  .. .. ..$ model.matrix 
  .. .. ..$ smooth.construct 
  ..$ y 
  .. ..$ num 
\end{Soutput}
\end{Schunk}
For writing a new estimation engine, the user can directly work with the \code{model.matrix}
elements, for linear effects, and the \code{smooth.construct} list, for smooth effects respectively.
The \code{smooth.construct} is a named list which is compiled using the \fct{smoothCon} function
of the \pkg{mgcv} package using the generic \fct{smooth.construct} method for setting up
smooth terms.
\begin{Schunk}
\begin{Sinput}
R> print(names(bf$x$mu$smooth.construct))
\end{Sinput}
\begin{Soutput}
[1] "s(x2)"       "s(x3)"       "te(lon,lat)"
\end{Soutput}
\end{Schunk}
In this example, the list contains three smooth term objects for parameter \code{mu} and
\code{sigma}.

As shown in Appendix~\ref{appendix:specialterms} the \fct{bamlss.frame} function can also
process special model terms, i.e., model terms that are not necessarily represented by
a linear matrix vector product.

\subsection{Family objects} \label{sec:families}

Family objects are important building blocks in the design of BAMLSS models.
They specify the distribution by collecting functions of the density,
respective log-likelihood, first-order derivatives of the log-likelihood w.r.t.\
predictors (the score function), and (optionally) second-order derivatives of
the log-likelihood w.r.t.\ predictors or their expectation (the Hessian).

The \pkg{bamlss} package can be easily extended by constructing families for specific
tasks, i.e., problems for which a likelihood can be formulated.
However, commonly used distributions are already implemented in \pkg{bamlss}; and the ones
from the \pkg{gamlss} package can also be accessed through the \pkg{bamlss} package.

We illustrate how to build a \pkg{bamlss} family by hand along the Gaussian
distribution, with density
$$
f(y\,|\,\mu,\sigma) = \frac{1}{\sqrt{2\pi}\sigma} \cdot \exp
\left( \frac{-(y-\mu)^2}{2\sigma^2} \right),
$$
and log-likelihood function
$$
\ell(\mu,\sigma\,|\,y) = - \frac{1}{2} \log(2\pi) - \log(\sigma) -
\frac{(y-\mu)^2}{2\sigma^2},
$$
for an individual observation. The sum of the log-likelihood function
over all observations is the target function of the optimization problem.

In the distributional regression framework the parameters are linked
to predictors by link functions,
$$
\mu = \eta_\mu, \qquad \log(\sigma) = \eta_\sigma.
$$
For the Gaussian $\mu$ and $\sigma$ are linked to $\eta_\mu$ and $\eta_\sigma$
by the identity function and the logarithm, respectively.

The score functions in \pkg{bamlss} are the first derivatives of the log-likelihood w.r.t.\
the predictors:
$$
s_\mu = \frac{\partial\ell}{\partial\eta_\mu} = \frac{\partial\ell}{\partial\mu} \cdot
\frac{\partial\mu}{\partial\eta_\mu} = \frac{y-\mu}{\sigma^2},
$$
and
$$
s_\sigma = \frac{\partial\ell}{\partial\eta_\sigma} = \frac{\partial\ell}{\partial\sigma} \cdot
\frac{\partial\sigma}{\partial\eta_\sigma} = -1 + \frac{(y-\mu)^2}{\sigma^2}.
$$
\begin{table}[ht!]
\centering
\begin{tabular}{lp{12cm}}
\hline
Name of element & Value \\ \hline
\code{family} & Character string with the name of the family. \\
\code{names} & Vector of character strings with the names of the parameters. \\
\code{links} & Vector of character strings with the names of the link functions \\
\code{d} & A function returning the density with arguments \code{y}, \code{par},
  \code{log = FALSE} (see below). \\
\code{score} & A list with functions (one for each parameter) returning the first derivatives of
  the log-likelihood w.r.t.\ predictors. \\
\code{hess} & A list with functions (one for each parameter) returning the negative second
  derivatives of the log-likelihood w.r.t.\ predictors. \\ \hline
\end{tabular}
\caption{\label{tab:families} Elements of the Gaussian distribution \code{"bamlss.family"} object.}
\end{table}
For the second derivative of the log-likelihood we are able to obtain the
negative expectation,
$$
\mathsf{E}(-\partial^2\ell / \partial\eta_{\mu}^{2} ) = \sigma^{-2},
$$
and
$$
\mathsf{E}(-\partial^2\ell / \partial\eta_{\sigma}^{2} ) = 2.
$$
Now we have to write a function that returns a \code{family.bamlss} object (S3)
which encapsulates functions for density, score and Hessian, and the names of
the family, parameter and link functions. The required elements are listed
in Table~\ref{tab:families}.

Merely all functions take as first argument the response \code{y} and as second
argument a named list holding the evaluated parameters \code{par} of the
distribution. The example implementation is shown in Appendix~\ref{appendix:families}.

Optionally, the \code{"family.bamlss"} object can be extended by functions for
\begin{itemize}
\item the cumulative distribution function \code{p(y, par, ...)},
\item the quantile function (the inverse cdf) \code{q(p, par)},
\item a random number generator \code{r(n, par)},
\item the log-likelihood \code{loglik(y, par)},
\item the expectation \code{mu(par, ...)},
\item initial values for optimization, which has to be a list containing a
  function for each parameter,
\item \code{...},
\end{itemize}
which can help to speed up optimization, or be convenient for
predictions and simulations.

For a list of all implemented families, please see the documentation of \code{?bamlss.family}.

\subsection{Estimation engines} \label{sec:engines}

Estimation engines in \pkg{bamlss} are usually based on the model frame setup function\linebreak
\fct{bamlss.frame} (see Section~\ref{sec:bf}), i.e., the functions all have a \code{x} argument,
which contains all the necessary model and penalty matrices, and a \code{y} argument, which is the
response (univariate or multivariate). In addition, an estimation engine usually has a \code{family}
argument, which specifies the model to be estimated. However, this is not a mandatory
argument, i.e., one could write an estimation function that is designed for one specific
problem, only.

The modeling setup is best explained by looking at the main estimation engines provided by
\pkg{bamlss}. The default optimizer using the \fct{bamlss} wrapper function is \fct{bfit}, which is a
backfitting routine. The most important arguments are
\begin{Code}
  bfit(x, y, family, start = NULL, weights = NULL, offset = NULL, ...)
\end{Code}
The default sampling engine in \pkg{bamlss} is \fct{GMCMC}, again the most important
arguments are
\begin{Code}
  GMCMC(x, y, family, start = NULL, weights = NULL, offset = NULL, ...)
\end{Code}
So basically, the arguments of the optimizer and the sampling function are the same, the
main difference is the return value. In \pkg{bamlss} optimizer functions usually return
a vector of estimated regression coefficients (parameters), while sampling functions 
return a matrix of parameter samples of class \code{"mcmc"} or \code{"mcmc.list"} (for details
see the  documentation of the \pkg{coda} package).

Internally, what the optimizer or sampling function is actually processing is not important
for the \code{bamlss()} wrapper function as long as a vector or matrix of parameters is
returned. For optimizer functions the return value needs to be named list with an element
\code{"parameters"}, the vector (also a matrix, e.g., for \fct{lasso} and \fct{boost} optimizers)
of estimated parameters. The most important requirement to make use of all extractor
functions like \fct{summary.bamlss}, \fct{predict.bamlss}, \fct{plot.bamlss},
\fct{residuals.bamlss}, etc., is to follow the naming convention of the returned estimates.
The parameter names are based on the names of the distribution parameters as specified in
the family object. For example, the family object \fct{gaussian\_bamlss} has parameter names
\code{"mu"} and \code{"sigma"}
\begin{Schunk}
\begin{Sinput}
R> gaussian_bamlss()$names
\end{Sinput}
\begin{Soutput}
[1] "mu"    "sigma"
\end{Soutput}
\end{Schunk}
Then, each distributional parameter can be modeled by parametric (linear) and
nonlinear smooth effect terms. The parametric part is indicated with \code{"p"} and the smooth part with
\code{"s"}. The names of the parametric coefficients are the names of the corresponding model
matrices as returned from \fct{bamlss.frame}. E.g., if two linear effects, with variables \code{"x1"}
and \code{"x2"}, enter the model for 
distributional parameter \code{"mu"}, then the final names are \code{"mu.p.x1"} and \code{"mu.p.x2"}.
Similarly for the smooth parts, if we model a variable \code{"x3"} using a regression spline as
provided by the \fct{s} function of the \pkg{mgcv} package, the
name is based on the names that are used by \fct{bamlss.frame} for the \fct{smooth.construct}
object. In this case the parameter names start with \code{"mu.s.s(x3)"}. If this smooth
term has 10 regression coefficients, then the final name must be
\begin{Schunk}
\begin{Sinput}
R> paste0("mu.s.s(x3)", ".b", 1:10)
\end{Sinput}
\begin{Soutput}
 [1] "mu.s.s(x3).b1"  "mu.s.s(x3).b2"  "mu.s.s(x3).b3" 
 [4] "mu.s.s(x3).b4"  "mu.s.s(x3).b5"  "mu.s.s(x3).b6" 
 [7] "mu.s.s(x3).b7"  "mu.s.s(x3).b8"  "mu.s.s(x3).b9" 
[10] "mu.s.s(x3).b10"
\end{Soutput}
\end{Schunk}
i.e., all smooth term parameters are named with \code{"b"} and a numerated.

An example of how to setup an estimation engine for \pkg{bamlss} for linear regression models is
given in Appendix~\ref{appendix:engine}. The example also provides details on the naming convention
and return values of optimizer and sampler functions.

\section{Flexible count regression for lightning reanalysis} \label{sec:application}

This section illustrates the workflow with \pkg{bamlss} along a small case study.
We want to build a statistical model linking positive counts of cloud-to-ground
lightning discharges to atmospheric quantities from a reanalysis dataset.

The region we focus on are the European Eastern Alps. Cloud-to-ground lightning
discharges---detected by the Austrian Lightning Detection and Information System
\citep[ALDIS, ][]{schulz2005}---are counted on grids with a mesh size of $32~km$.
The lightning observations are available for the period 2010--2018.
The reanalysis data comes from the fifth generation of the ECMWF (European
Centre for Medium-Range Weather Forecasts) atmosphheric reanalyses of the
global climate \citep{era5}. ERA5 provides a globally complete and consistent
pseudo-observations of the atmosphere using the laws of physics. The horizontal
resolution is approx.\ $32~km$, while the temporal resolution is hourly and covers the years
from 1979 to present.
In this example application we work only with a small subset of the data, which
can be assessed from the accompanying \proglang{R} package \pkg{FlashAustria}
\citep{FlashAustria}. The data is loaded with
\begin{Schunk}
\begin{Sinput}
R> data("FlashAustria", package = "FlashAustria")
R> head(FlashAustriaTrain)
\end{Sinput}
\begin{Soutput}
  counts      d2m   q_prof_PC1 cswc_prof_PC4 t_prof_PC1 v_prof_PC2
1      2 291.3184 -0.011472293  7.168725e-06  15.922548  2.5646172
2     16 283.5004  0.001007288  1.612870e-05  -9.758380  0.7955608
3      1 291.0506 -0.005590341 -3.226052e-06  20.274007  7.5535312
4      7 288.0358 -0.006293043  3.715074e-05  14.258116  5.8523424
5     41 288.4433 -0.006315605  3.509800e-05   8.757239  8.3675943
6      1 286.6035 -0.001597900 -3.195042e-06  -3.433136 -3.4291366
  sqrt_cape   sqrt_lsp
1  45.37480 0.00000000
2  14.62869 0.00350679
3  20.31514 0.00000000
4  12.26630 0.00000000
5  20.18042 0.00000000
6  10.63068 0.00000000
\end{Soutput}
\begin{Sinput}
R> nrow(FlashAustriaTrain)
\end{Sinput}
\begin{Soutput}
[1] 12000
\end{Soutput}
\end{Schunk}
The motivation for this application is as follows: Lightning counts are not
modeled within the atmospheric reanalyses. Lightning observations are only
available for the period 2010--2018. With a statistical model on hand one could
predict lightning counts for the time before 2010 and thus analyze lightning
events in the past for which no observations are available.

The response of our statistical model are positive counts, with a mean of
13.61, and a variance of
1180.63. Thus, we are facing a
truncated count data distribution which is highly overdispersive.  In order to
capture the truncation of the data and its overdispersion we employ a
zero-truncated negative binomial distribution \citep{cameron2013count}, which
is specified by two parameters $\mu > 0$ and $\theta > 0$. $\mu$ is the
expectation of the underlying untruncated negative binomial, and $\theta$
modifies the variance of the untruncated negative binomial by
$\mathrm{VAR}(\tilde{Y}) = \mu + \mu^2/\theta$, where $\tilde{Y}$ is a latent
random variable following the underlying untruncated negative binomial
distribution.

\begin{figure}[t!]
\centering
\includegraphics{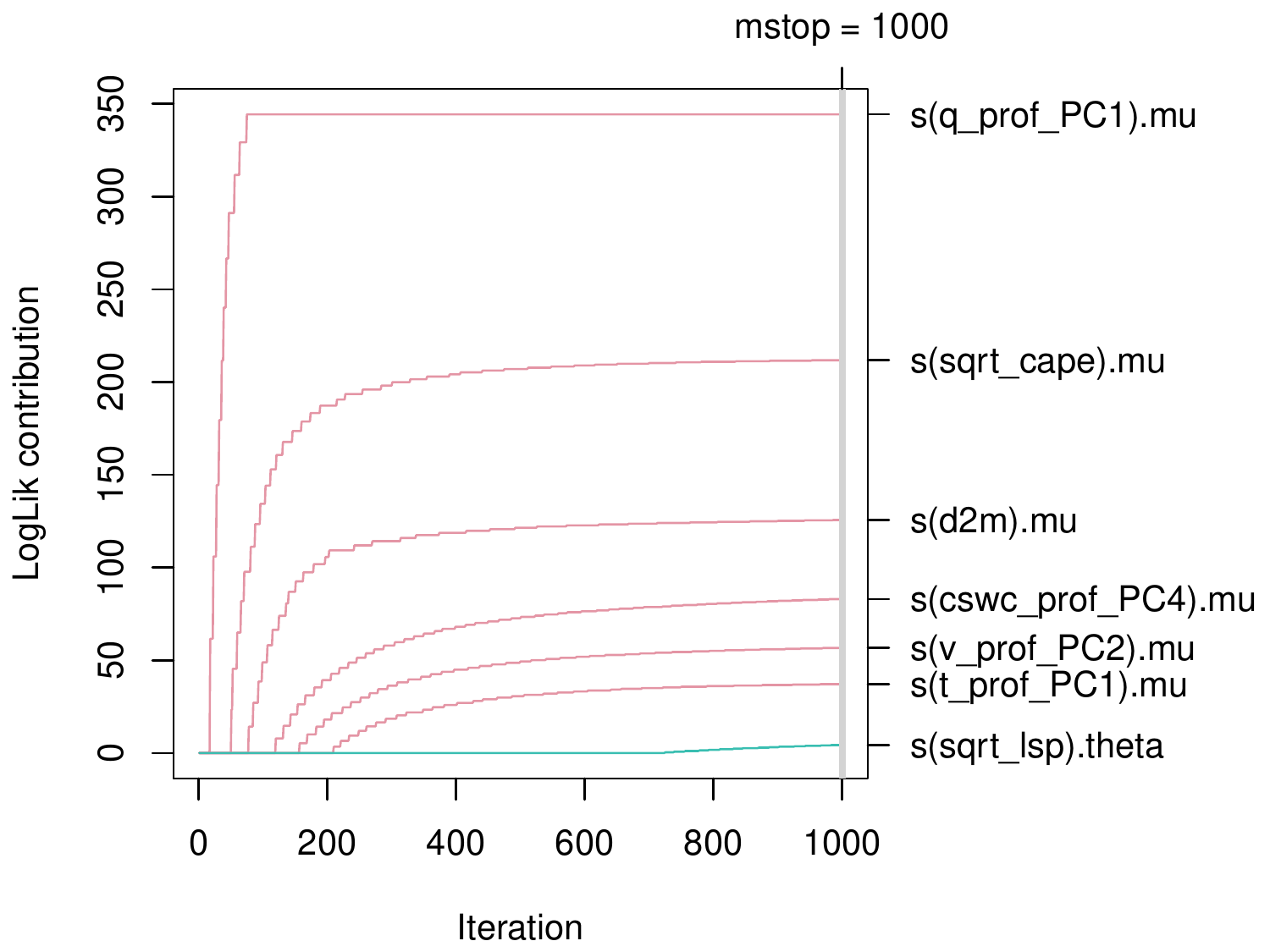}
\caption{Contribution to the log-likelihood of individual terms during
gradient boosting.}
\label{fig:appboost}
\end{figure}

The zero-truncated negative binomial distribution is implemented as
\code{ztnbinom_bamlss()} within \pkg{bamlss}.  In order to specify smooth terms
form both distributional parameter, the formula has to be a \code{list}.
The first element specifies terms for the response \code{counts}, which is named $\mu$ in the
\pkg{bamlss} familiy object.
The second element specifies the formula for parameter $\theta$. Hence well known
for their sampling properties, we are
applying P-splines \citep{bamlss:Eilers+Marx:1996} for all terms. Specifying
smooth terms within \pkg{bamlss} formulae builds on the \pkg{mgcv} infrastructure
\citep{bamlss:Wood:2019} provided by \code{s()}, which leads to the following
specification of the model (formula):
\begin{Schunk}
\begin{Sinput}
R> f <- list(
+    counts ~ s(d2m, bs = "ps") + s(q_prof_PC1, bs = "ps") +
+      s(cswc_prof_PC4, bs = "ps") + s(t_prof_PC1, bs = "ps") +
+      s(v_prof_PC2, bs = "ps") + s(sqrt_cape, bs = "ps"), 
+    theta ~ s(sqrt_lsp, bs = "ps")
+  )
\end{Sinput}
\end{Schunk}
Now we have all ingredients on hand to feed the standard interface for
statistical models in \proglang{R}: A formula \code{f}, a family
\code{ztnbinom_bamlss()}, and a data set \code{FlashAustriaTrain}. Within the
\code{bamlss()} call we also provide arguments which are passed forward to the
optimizer and the sampler. We choose the gradient boosting optimizer
\code{boost()} in order to find initial values for the default sampler
\code{GMCMC()}. Gradient boosting proved to offer a very stable method for
finding regression coefficients that serve as initial values for a MCMC sampler
\citep{bamlss:Simon+Mayr+Umlauf+Zeileis:2019}.  We set the number of iteration
to $1000$. For the sampling we allow another $1000$ iterations as burn-in
phase, and apply a thinning of the resulting chain of $5$.  Running
\code{n.iter = 6000} iterations in total leads to $1000$ MCMC samples in the
end:
\begin{Schunk}
\begin{Sinput}
R> set.seed(111)
R> b <- bamlss(f, family = "ztnbinom", data = FlashAustriaTrain,
+    optimizer = boost, maxit = 1000,         ## Boosting arguments.
+    thin = 5, burnin = 1000, n.iter = 6000)  ## Sampler arguments.
\end{Sinput}
\begin{Soutput}
logLik -36930.0 eps 0.0003 iteration 1000 qsel 7
elapsed time:  5.31min
Starting the sampler...
|********************| 100%  0.00sec 27.76min
\end{Soutput}
\end{Schunk}
The model was fitted on a single core Intel i7-7700 CPU with 3.60GHz and 16~GB memory,
on which the boosting took about 5.3~minutes and the MCMC sampling took about
27.8~minutes. As a first diagnostic we check the log-likelihood contributions of the
individual terms during the boosting optimization (Figure~\ref{fig:appboost}).
\begin{Schunk}
\begin{Sinput}
R> pathplot(b, which = "loglik.contrib", intercept = FALSE)
\end{Sinput}
\end{Schunk}
\setkeys{Gin}{width=.7\textwidth}
After 1000~iterations the term \code{s(q_prof_PC1).mu} has the highest contribution
to the log-likelihood with %
344 %
followed by \code{s(sqrt_cape).mu} with %
212.
The term of the parameter $\theta$ \code{s(sqrt_lsp).theta} has a relatively small
contribution with %
4.
The overall message of this diagnostic is that the contributions to the
log-likelihood at the end of the boosting procedure are very small and that
the algorithm approached a stable state, which suggest that we retrieve
reasonable initial values for the MCMC sampling.

\begin{figure}[t!]
\centering
\includegraphics{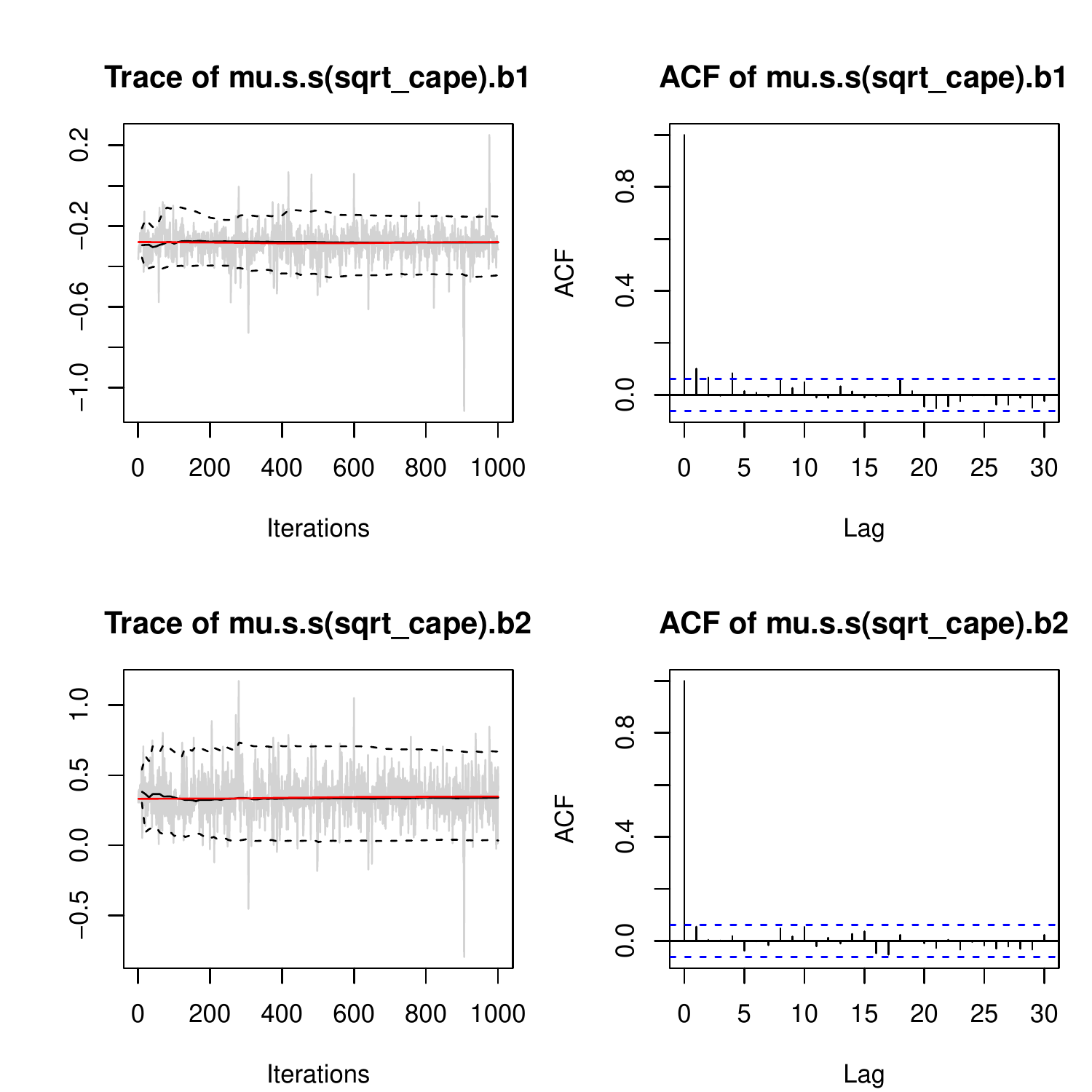}
\caption{MCMC trace (left panels) and auto-correlation (right panels) for two
splines from the term \code{s(sqrt\_cape)} of the model \code{mu}.}
\label{fig:apptrace}
\end{figure}

The MCMC chains are investigated by looking directly at the traces of the chains
and with the auto-correlation function of the chains.
\begin{Schunk}
\begin{Sinput}
R> plot(b, model = "mu", term = "s(sqrt_cape)", which = "samples")
\end{Sinput}
\end{Schunk}
Figure \ref{fig:apptrace} shows the traces and the auto-correlation functions
for two regression coefficients of the term \code{s(sqrt_cape)}. The traces
reveal samples around stables means. This suggests that the 1000 boosting
iterations and the 1000 burn-in samples were sufficient in order to approach
reasonable starting values for the sampling. The auto-correlation functions
reveal that after the thinning hardly any correlation remains between
consecutive samples.

\begin{figure}[t!]
\centering
\setkeys{Gin}{width=1\textwidth}
\includegraphics{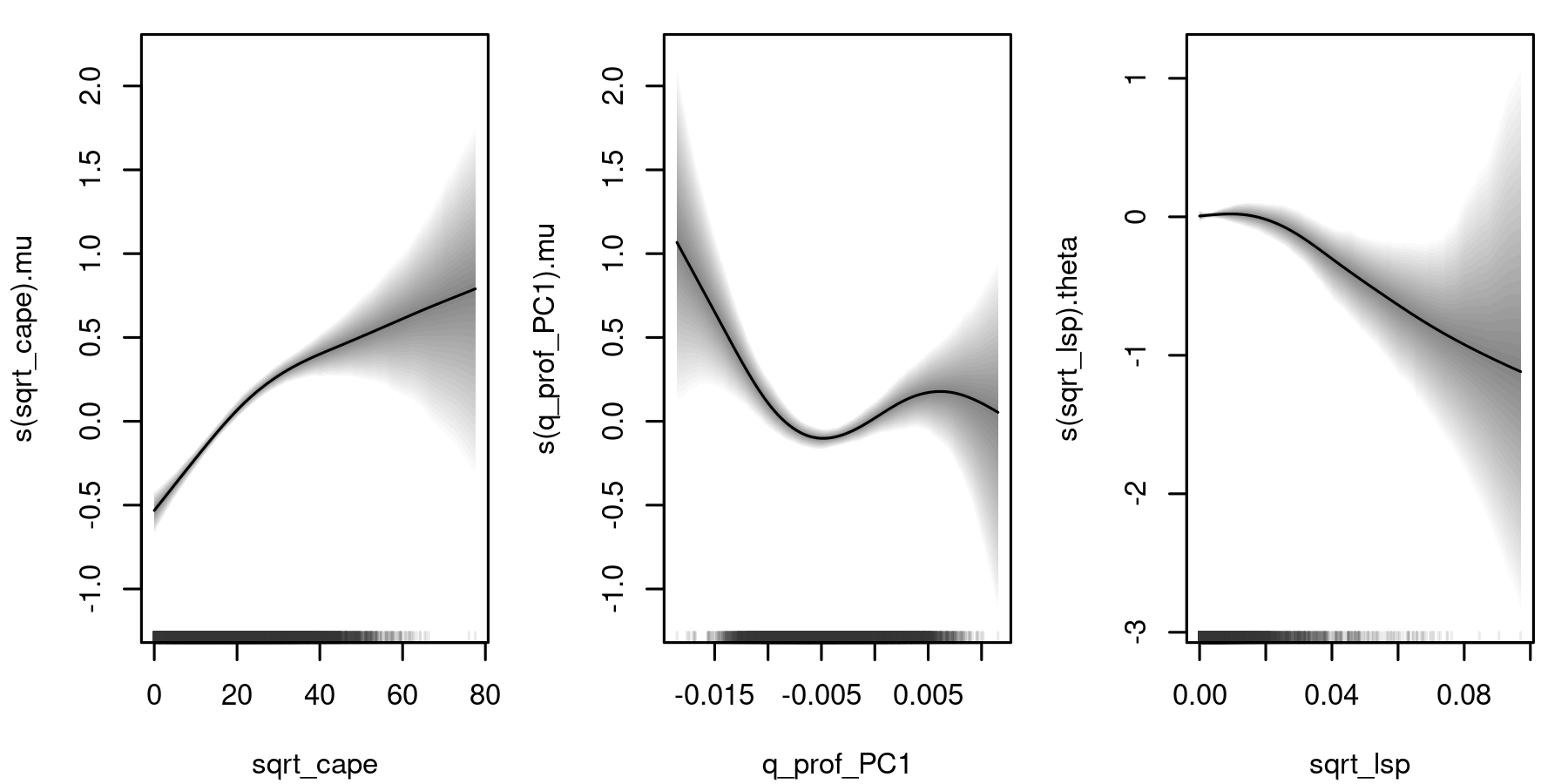}
\caption{Effect of the terms \code{s(sqrt\_cape)} and \code{s(q\_prof\_PC1)}
from the model \code{mu} and term \code{s(sqrt\_lsp)} from model \code{theta}.
Credible intervals derived from MCMC samples.}
\label{fig:appeffect}
\end{figure}

As these diagnostics suggest that a reasonable initial state for the sampling
has been found and the samples are independent draws from the posterior, one
can go further and investigate the estimated effects. The boosting summary
(Figure~\ref{fig:appboost}) revealed that the terms \code{s(sqrt_cape)} and
\code{s(q_prof_PC1)} had a large contribution for improving the fit. 
Looking at these effects illustrate how the atmospheric parameters
of the reanalyses are related to lightning events (Figure~\ref{fig:appeffect}),
and thus help to understand the physics associated with lightning events.
The effects are presented on the scale of the linear predictor, i.e., the
log scale.
\begin{Schunk}
\begin{Sinput}
R> plot(b, term = c("s(sqrt_cape)", "s(q_prof_PC1)", "s(sqrt_lsp)"))
\end{Sinput}
\end{Schunk}
\code{s(sqrt_cape)} reveals a monotonic increasing shape. In the
range from $0$--$30$ the effect increases linearly with small credible intervals.
For higher values the effect flattens and shows large credible intervals which are
associated with the small amount of data in that range. Physically the shape of the
effect is meaningful as more convective available potential energy has the
potential to lead to heavier lightning events.
\code{s(q_prof_PC1)} reveals areas of large credible intervals at the left and
right bounds of the range due to small amount of data. In the mid-range an
increasing effect is identified. As \code{q_prof_PC1} is the leading principal
component of the vertical profile of specific humidity, one has to consider the
corresponding spatial mode (not shown) for interpretation: Higher values of
\code{q_prof_PC1} are linked to more moisture in the lower atmosphere,
which is also available as a source of \emph{latent energy}, i.e., energy that
becomes free when water transfers from the gas to the liquid phase.

Finally it is interesting to look at the effect acting on the link scale of
the parameter $\theta$, \code{s(sqrt_lsp)} (right panel in Figure~\ref{fig:appeffect}).
\code{sqrt_lsp} is the square root of large scale precipitation, i.e.,
precipitation that is not linked to convective processes and thus it is
not related to strong lightning events. The effect shows following relationship:
Higher values of \code{sqrt_lsp} lead to smaller $\theta$, which increases the
variance of the distribution.

Before applying the model, i.e., predicting lightning cases before 2010, we
check the marginal calibration of the distribution by hanging rootogram, a tool
popular for the evaluation of count data regression models
\citep{kleiber2016visualizing}. First we predict the distributional parameter
on out-of-sample data \code{FlashAustriaEval} for which lightning
observations are on hand
\begin{Schunk}
\begin{Sinput}
R> fit <- predict(b, newdata = FlashAustriaEval, type = "parameter")
R> str(fit)
\end{Sinput}
\begin{Soutput}
List of 2
 $ mu   : num [1:6000] 0.0159 0.0328 0.0126 0.0265 0.0458 ...
 $ theta: num [1:6000] 0.000706 0.000707 0.000712 0.000709 0.000704 ...
\end{Soutput}
\end{Schunk}
\code{predict()} returns a \code{list}, of which each element is named
as a distributional parameter and contains by default a vector of
predictions. Each prediction is the average of the predictions obtained
by all MCMC samples. The resulting \code{list} can be used to
derive further quantities by employing the functions of the \pkg{bamlss}
family that can be extracted using \code{family()},
\begin{Schunk}
\begin{Sinput}
R> fam <- family(b)
R> fam
\end{Sinput}
\begin{Soutput}
Family: ztnbinom 
Link function: mu = log, theta = log
---
Derivative functions:
 ..$ score
 .. ..$ mu
 .. ..$ theta
 ..$ hess
 .. ..$ mu
 .. ..$ theta
\end{Soutput}
\end{Schunk}
The family contains functions to map the predictors to the parameter
scale, density, cumulative distribution function, log-likelihood, and
scores and Hessian. We apply the density to compute the expected
frequencies of the positive counts. The function \code{...$d()} takes the
quantile as first argument, and the \code{list} with the parameters,
as returned by \code{predict()}, as a second argument. The expected frequencies can then be computed
by
\begin{Schunk}
\begin{Sinput}
R> expect <- sapply(1:50, function(j) sum(fam$d(j, fit)))
\end{Sinput}
\end{Schunk}
In order to plot the rootogram, we have to name the vector and
coerce it to an object of class \code{table}. The verifying
observed frequencies can be directly obtained by \code{table}.
\begin{Schunk}
\begin{Sinput}
R> names(expect) <- 1:50
R> expect <- as.table(expect)
R> obsrvd <- table(FlashAustriaEval$counts)[1:50]
\end{Sinput}
\end{Schunk}

\begin{figure}[p!]
\setkeys{Gin}{width=.7\textwidth}
\centering
\includegraphics{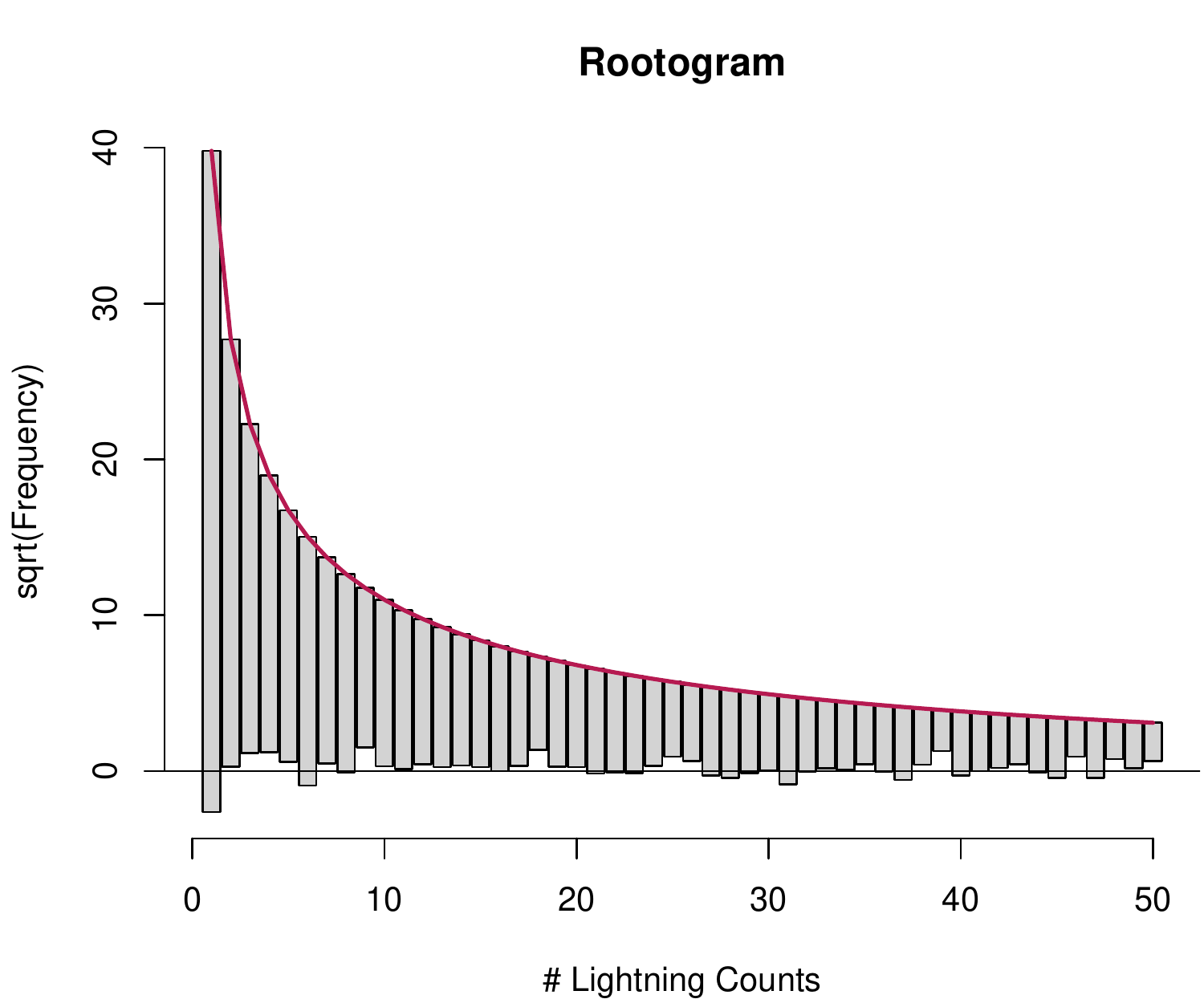}
\caption{Hanging rootogram for evaluating calibration of count data model
on out-of-sample data. Red line indicates the expected frequencies on the
square root scale. Gray bars indicate observed frequencies on square root
scale hanging from the red line.}
\label{fig:approoto}
\end{figure}

\begin{figure}[p!]
\setkeys{Gin}{width=0.8\textwidth}
\centering
\includegraphics{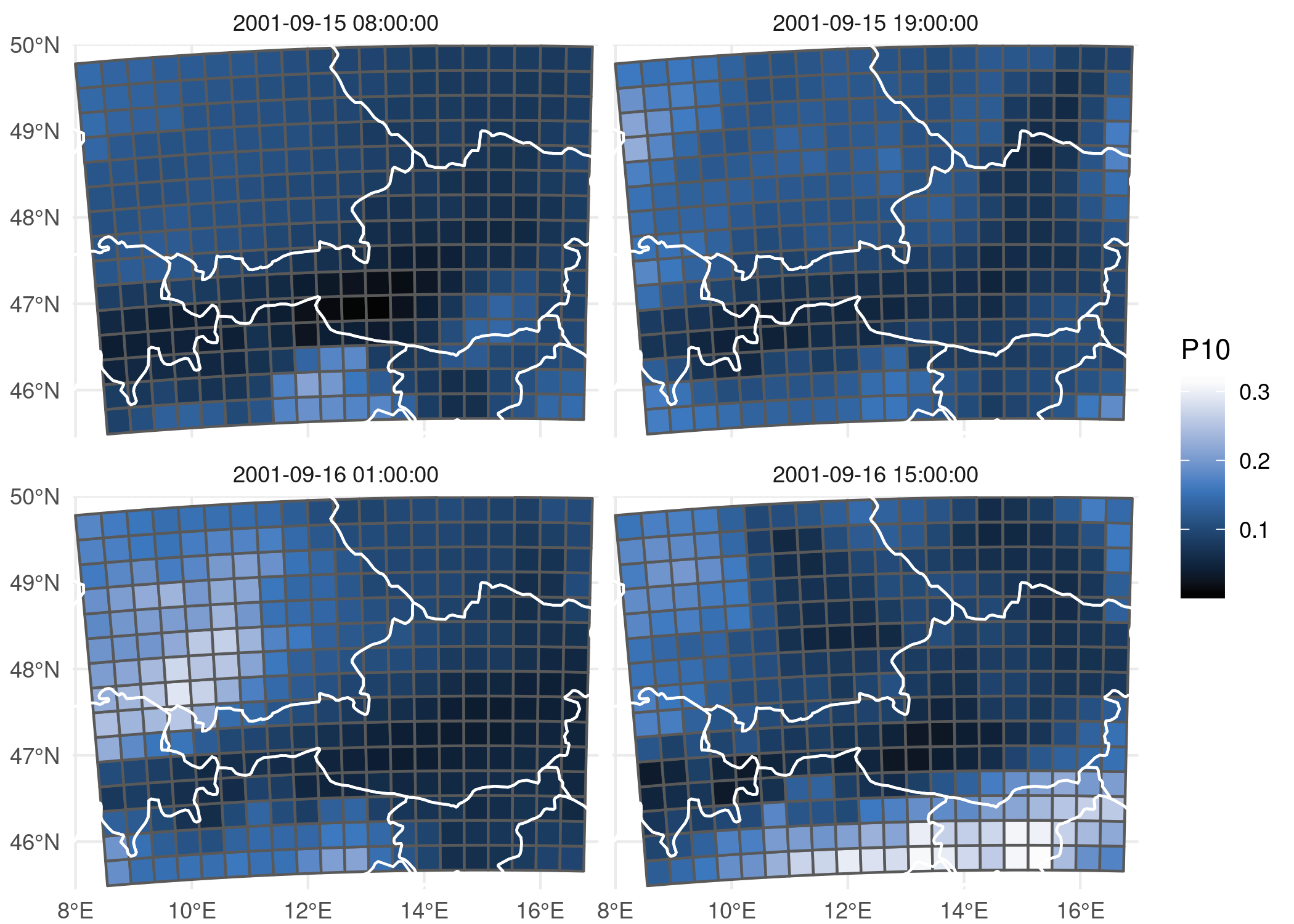}
\caption{\label{fig:appcase} A probabilistic reconstruction of lightning counts occured on
September 15 2001 at 6~UTC, 17~UTC and 23~UTC and on September 16 2001 at
13~UTC, i.e., the probability of having observed $10$ or more counts within one
grid box.}
\end{figure}

The observed and expected frequencies can be plugged into the
default method of \code{rootogram} from the \pkg{countreg}
package \citep{zeileis2008regression}.
\begin{Schunk}
\begin{Sinput}
R> library("countreg")
R> rootogram(obsrvd, expect, xlab = "# Lightning Counts", main = "Rootogram")
\end{Sinput}
\end{Schunk}
The rootogram reveals reasonable calibration of the model though it is slightly
underestimating the number of events with a single lightning discharge.
Now given good convergence and sample characteristics of the gradient
boosting optimizer and MCMC sampler, physically interpretable effects, and
good out-of-sample calibration, we can take our model and predict a case
for the period before 2010, for which no lightning data are available.
The case of interest is a front moving from the West to the East on the
Northern side of the Alps on 2001-09-15 and 2001-09-16.
The case data \code{FlashAustriaCase} contains additional columns containing
time and space information, and is of class \code{sf} \citep{pebesma2018sf}.
We predict the parameters for this case, and derive the probability of
observing 10 or more flashes within a grid box conditioned on a thunderstorm activity,
by applying the cumulative distribution function \code{...$p} of the family
\begin{Schunk}
\begin{Sinput}
R> library("sf")
R> fit <- predict(b, newdata = FlashAustriaCase, type = "parameter")
R> FlashAustriaCase$P10 <- 1 - fam$p(9, fit)
\end{Sinput}
\end{Schunk}
We visualize this case by employing \code{ggplot()} \citep{wickham2016gg},
and the Oslo color scale from the \pkg{colorspace} package
\citep{zeileis2019colorspace}. The country borders \code{world} are retrieved
from the \pkg{rnaturalearth} package \citep{south2017rnaturalearth}. 
\begin{Schunk}
\begin{Sinput}
R> library("ggplot2")
R> world <- rnaturalearth::ne_countries(scale = "medium", returnclass = "sf")
R> ggplot() + geom_sf(aes(fill = P10), data = FlashAustriaCase) +
+    scale_fill_continuous_sequential("Oslo", rev = TRUE) +
+    geom_sf(data = world, col = "white", fill = NA) +
+    coord_sf(xlim = c(7.95, 17), ylim = c(45.45, 50), expand = FALSE) +
+    facet_wrap(~time) + theme_minimal()
\end{Sinput}
\end{Schunk}
The maps are shown in Figure~\ref{fig:appcase} and reveal that the probability for strong
lightning events increases during 2001-09-15 between 6 and 17~UTC. During night time the front
occurs, which can be nicely seen at 23~UTC. The propagation of the front is blocked
by the main Alpine ridge located at $47^\circ~N$. On the subsequent day
2001-09-16 one can see that the probability on the downwind side of the
Alps has increased.

\section*{Acknowledgments}

Thorsten Simon acknowledges the funding by the Austrian Science Fund (FWF, grant no.~P31836)

\bibliography{bamlss}

\pagebreak

\begin{appendix}

\section[Custom CRPS() function]{Custom \fct{CRPS} function}
\label{appendix:crps}

The \proglang{R} package \pkg{scoringRules} \citep{bamlss:Jordan+Krueger+Lerch:2018} provides
tools for model calibration checks. A commonly used measure is the CRPS. Since the number of
candidate distributions in BAMLSS is quite large, it can happen that the CRPS for some
distributions is not implemented. In such a case the reader can implement the CRPS using
numerical integration. The following \proglang{R} code implements the \fct{CRPS} to be used with
\pkg{bamlss} and a numeric response, e.g., which can be used with the motorcycle accident model
presented in Section~\ref{sec:locscalemodel}.
\begin{Schunk}
\begin{Sinput}
R> CRPS <- function(object, newdata = NULL) {
+    yname <- response_name(object)
+    fam <- family(object)
+    if(is.null(fam$p))
+      stop("no p() function in family object!")
+    if(is.null(newdata))
+      newdata <- model.frame(object)
+    n <- nrow(newdata)
+    crps <- rep(0, n)
+    par <- as.data.frame(predict(object, newdata = newdata, type = "parameter"))
+    for(i in 1:n) {
+      foo <- function(y) {
+        (fam$p(y, par[i, , drop = FALSE]) - 1 * (y >= newdata[[yname]][i]))^2
+      }
+      crps[i] <- integrate(foo, -Inf, Inf)$value
+    }
+    return(crps)
+  }
\end{Sinput}
\end{Schunk}

\section[Gaussian family object]{Gaussian family object} \label{appendix:families}

The following \proglang{R} code shows an example implementation of the Gaussian distribution
as presented in Section~\ref{sec:families}.
\begin{Schunk}
\begin{Sinput}
R> Gauss_bamlss <- function(...) {
+    f <- list(
+      "family" = "mygauss",
+      "names"  = c("mu", "sigma"),
+      "links"  = c(mu = "identity", sigma = "log"),
+      "d" = function(y, par, log = FALSE) {
+        dnorm(y, mean = par$mu, sd = par$sigma, log = log)
+      },
+      "p" = function(y, par, ...) {
+        pnorm(y, mean = par$mu, sd = par$sigma, ...)
+      },
+      "r" = function(n, par) {
+        rnorm(n, mean = par$mu, sd = par$sigma)
+      },
+      "q" = function(p, par) {
+        qnorm(p, mean = par$mu, sd = par$sigma)
+      },
+      "score" = list(
+        mu = function(y, par, ...) {
+          drop((y - par$mu) / (par$sigma^2))
+        },
+        sigma = function(y, par, ...) {
+          drop(-1 + (y - par$mu)^2 / (par$sigma^2))
+        }
+      ),
+      "hess" = list(
+        mu = function(y, par, ...) {
+          drop(1 / (par$sigma^2))
+        },
+        sigma = function(y, par, ...) { 
+          rep(2, length(y))
+        }
+      )
+    )
+    class(f) <- "family.bamlss"
+    return(f)
+  }
\end{Sinput}
\end{Schunk}

\section[Special model terms]{Special model terms} \label{appendix:specialterms}

The default estimation engines \fct{bfit} and \fct{GMCMC} (also the gradient boosting optimizer function
\fct{boost}) in \pkg{bamlss} provide support for the implementation of special model terms, i.e.,
model terms that cannot be represented by the \pkg{mgcv} smooth term constructor infrastructures.
One simple example of such a special model term is a nonlinear growth curve, e.g., a nonlinear
Gompertz curve
$$
f(x; \boldsymbol{\beta}) = \beta_1 \cdot \exp(-\beta_2 \cdot \exp(-\beta_3 \cdot x)),
$$
but also the lasso model term constructor \fct{la} presented in Section~\ref{sec:flexm} is a
special \pkg{bamlss} model term. The special model term constructor is needed in this case, since
the growth curve is nonlinear in the parameters $\boldsymbol{\beta}$, hence, the default backfitting
and sampling strategies cannot be applied. Fortunately, estimation algorithms in distributional
regression can be split into separate updating equations (see also Section~\ref{sec:algos}). This
means that each model term can have its own updating function. The user interested in this feature
only needs to write a new \fct{smooth.construct} and \fct{Predict.matrix} method.

The following \proglang{R} code implements a Gompertz growth model term
which can be used by the default optimizer function \fct{bfit} and sampling function \fct{GMCMC}
of the \pkg{bamlss} package. The new \fct{smooth.construct} method is
\begin{Schunk}
\begin{Sinput}
R> smooth.construct.gc.smooth.spec <- function(object, data, knots) 
+  {
+    object$X <- matrix(as.numeric(data[[object$term]]), ncol = 1)
+    center <- if(!is.null(object$xt$center)) {
+      object$xt$center
+    } else TRUE
+    object$by.done <- TRUE
+    if(object$by != "NA")
+      stop("by variables not supported!")
+  
+    ## Begin special elements to be used with bfit() and GMCMC().
+    object$fit.fun <- function(X, b, ...) {
+      f <- b[1] * exp(-b[2] * exp(-b[3] * drop(X)))
+      if(center)
+        f <- f - mean(f)
+      f
+    }
+    object$update <- bfit_optim
+    object$propose <- GMCMC_slice
+    object$prior <- function(b) { sum(dnorm(b, sd = 1000, log = TRUE)) }
+    object$fixed <- TRUE
+    object$state$parameters <- c("b1" = 0, "b2" = 0.5, "b3" = 0.1)
+    object$state$fitted.values <- rep(0, length(object$X))
+    object$state$edf <- 3
+    object$special.npar <- 3 ## Important!
+    ## End special elements.
+  
+    ## Important, This is a special smooth constructor!
+    class(object) <- c("gc.smooth", "no.mgcv", "special")
+  
+    object
+  }
\end{Sinput}
\end{Schunk}
In principle, the setup is very similar to the smooth constructor functions provided by the
\pkg{mgcv} package. Only few elements need to be added:
\begin{itemize}
\item \fct{fit.fun}: A function of the data \code{X} and parameter vector \code{b} that
  evaluates the fitted values.
\item \fct{update}: An updating function to be used with optimizer \fct{bfit}.
\item \fct{propose}: A MCMC propose function to be used with sampler \fct{GMCMC}.
\item \fct{prior}: Function of the parameters \code{b} that evaluates the log-prior. Note,
  additional functions can be \fct{grad} and \code{hess} that evaluate the first and second
  derivative of the log-prior w.r.t.\ the parameters \code{b}.
\item \code{fixed}: Is the number of degrees of freedom fixed or not?
\item \code{state}: This is a named list with starting values for the \code{"parameters"},
  the \code{"fitted.values"} and degrees of freedom \code{"edf"}. Note that regression
  coefficients are always named with \code{"b*"} and shrinkage or smoothing variances with
  \code{"tau2*"} in the \code{"parameters"} vector.
\item \code{special.npar}: How many parameters does this model term have in total? This is needed
  for internal setup, because the Gompertz function has three parameters but the design matrix only one
  column.
\end{itemize}
To compute predictions of this model term a new method for the \fct{Predict.matrix} function needs
to be implemented, too.
\begin{Schunk}
\begin{Sinput}
R> Predict.matrix.gc.smooth <- function(object, data, knots) 
+  {
+    X <- matrix(as.numeric(data[[object$term]]), ncol = 1)
+    X
+  }
\end{Sinput}
\end{Schunk}

Special model terms can then be used with the constructor function \fct{s2}. To illustrate the
this feature in \pkg{bamlss}, we simulate heteroskedastic growth data with
$$
\texttt{y} \sim \mathcal{N}(\mu = 2 + 1 / (1 + \exp(0.5 \cdot (15 - \texttt{time}))),
  \log(\sigma) = -3 + 2 \cdot \cos(\texttt{time}/30 \cdot 6 - 3))
$$
and subsequently estimate the model with slice sampling \citep{bamlss:Neal:2003} for $\boldsymbol{\beta}$ in the
MCMC algorithm using the following \proglang{R} code
\begin{figure}[!t]
\centering
\setkeys{Gin}{width=0.4\textwidth}
\includegraphics{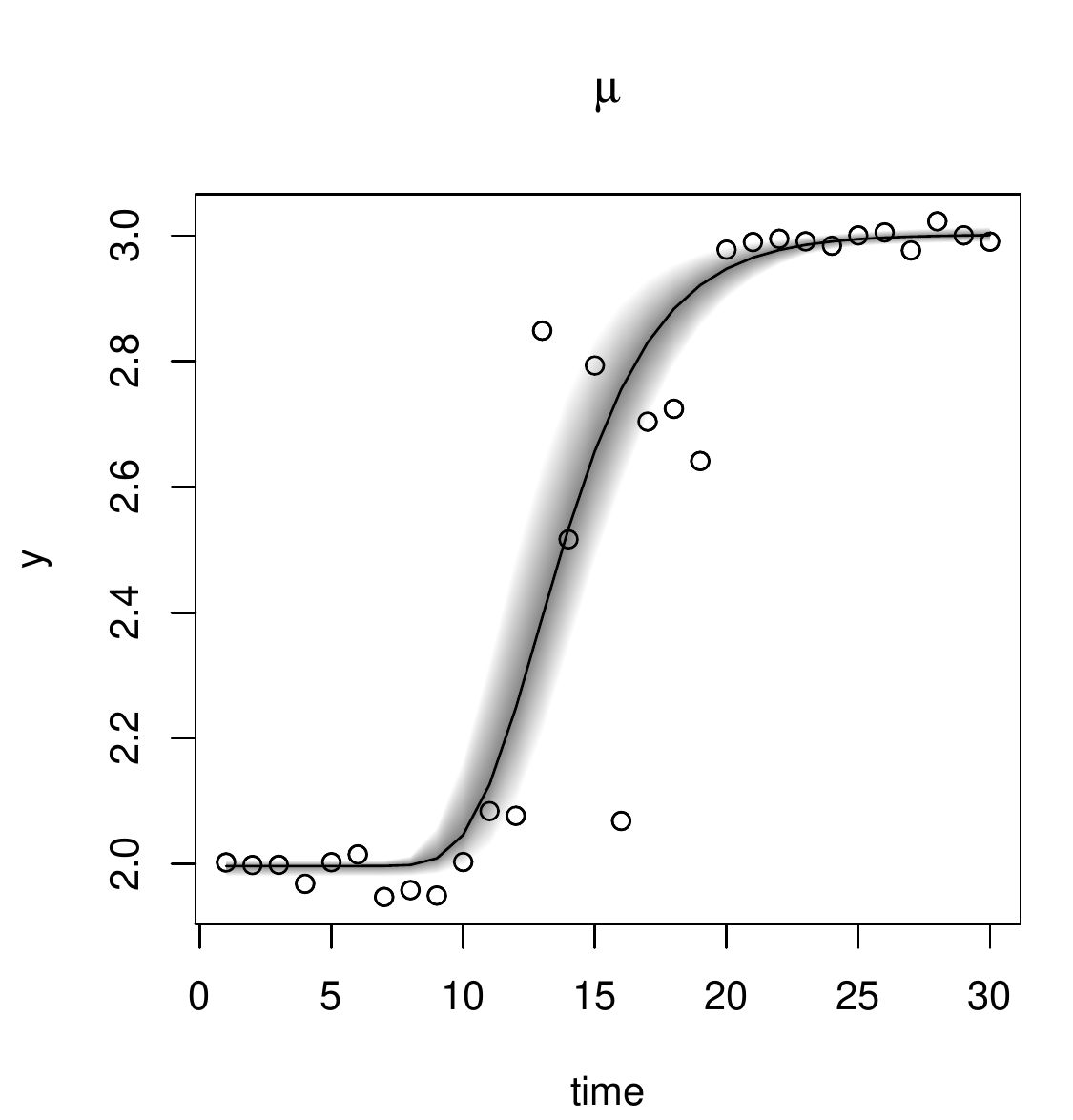}
\includegraphics{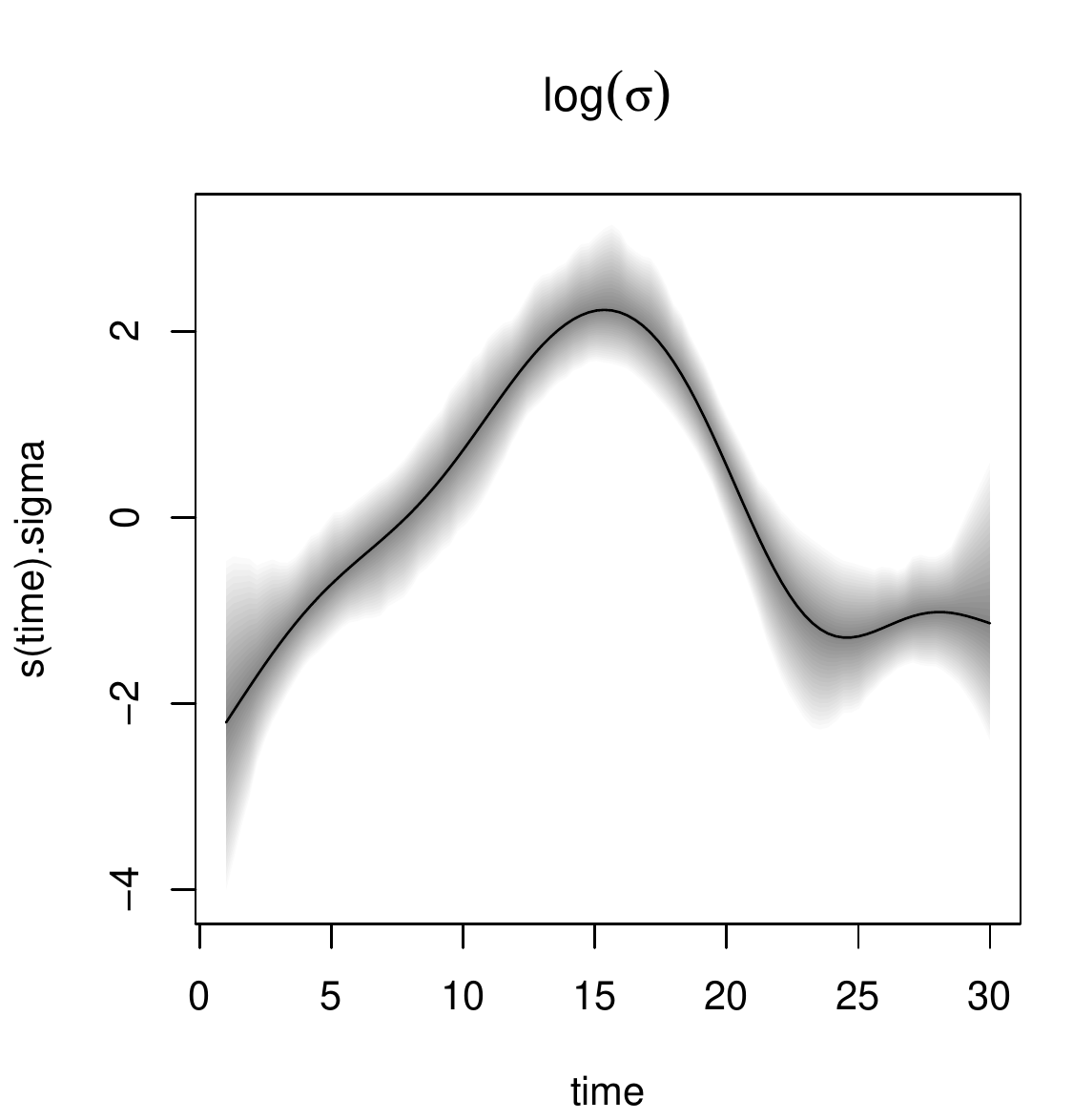}
\caption{\label{fig:gcurve} Estimated nonlinear effects on parameter $\mu$ and $\sigma$ of the
  simulated growth curve example. Gray shaded areas represent 95\% credible intervals.}
\end{figure}
\begin{Schunk}
\begin{Sinput}
R> set.seed(111)
R> d <- data.frame("time" = 1:30)
R> d$y <- 2 + 1 / (1 + exp(0.5 * (15 - d$time))) +
+    rnorm(30, sd = exp(-3 + 2 * cos(d$time/30 * 6 - 3)))
R> f <- list(
+    y ~ s2(time, bs = "gc"),
+    sigma ~ s(time)
+  )
R> b <- bamlss(f, data = d, optimizer = bfit, sampler = GMCMC)
R> plot(b)
\end{Sinput}
\end{Schunk}
The estimated effects are shown in Figure~\ref{fig:gcurve}.
The growth curve mean function estimate seems to fit the data quite well. Also, the nonlinear
relationship for parameter $\sigma$ could be captured by the model.

In summary, in order to build up special \pkg{bamlss} model terms only a few things have to be
considered. The example \proglang{R} code for the Gompertz smooth constructor given here is a good
starting point for readers interested in using this feature.

\section[Engines for linear regression]{Model fitting engines for linear regression} \label{appendix:engine}

In the following, to explain the setup and the naming convention of estimation engines in
more detail, we implement
\begin{itemize}
\item a new family object for simple linear models $y = x^{\top}\boldsymbol{\beta} + \varepsilon$ with
  $\varepsilon \sim N(0, \sigma^2)$,
\item and set up an optimizer function,
\item and additionally a MCMC sampling function.
\end{itemize}
For illustration, the family object is kept very simple, we only model the mean function 
in terms of covariates.
\begin{Schunk}
\begin{Sinput}
R> lm_bamlss <- function(...) {
+    f <- list(
+      "family" = "LM",
+      "names" = "mu",
+      "links" = "identity",
+      "d" = function(y, par, log = FALSE) {
+        sigma <- sqrt(sum((y - par$mu)^2) / (length(y) - .lm_bamlss.p))
+        dnorm(y, mean = par$mu, sd = sigma, log = log)
+      },
+      "p" = function(y, par, ...) {
+        sigma <- sqrt(sum((y - par$mu)^2) / (length(y) - .lm_bamlss.p))
+        pnorm(y, mean = par$mu, sd = sigma, ...)
+      }
+    )
+    class(f) <- "family.bamlss"
+    return(f)
+  }
\end{Sinput}
\end{Schunk}
Now, for setting up the estimation functions we first simulate some data using the
\fct{GAMart} function, afterwards the necessary \code{"bamlss.frame"} can be created with
\begin{Schunk}
\begin{Sinput}
R> d <- GAMart()
R> bf <- bamlss.frame(num ~ x1 + x2, data = d, family = "lm")
R> print(bf)
\end{Sinput}
\begin{Soutput}
'bamlss.frame' structure: 
  ..$ call 
  ..$ model.frame 
  ..$ formula 
  ..$ family 
  ..$ terms 
  ..$ x 
  .. ..$ mu 
  .. .. ..$ formula 
  .. .. ..$ fake.formula 
  .. .. ..$ terms 
  .. .. ..$ model.matrix 
  ..$ y 
  .. ..$ num 
\end{Soutput}
\end{Schunk}
As noted above, the object is a named list with elements \code{"x"} and \code{"y"}, which will be
passed to the estimation functions. For the moment, since we only implement a linear 
model, we need to work with the linear model matrix that is part of the \code{bf} object.
\begin{Schunk}
\begin{Sinput}
R> head(bf$x$mu$model.matrix)
\end{Sinput}
\begin{Soutput}
  (Intercept)        x1         x2
1           1 0.2905102 0.32659717
2           1 0.5090036 0.03047384
3           1 0.3900498 0.82453055
4           1 0.3650458 0.24858952
5           1 0.5219909 0.19089833
6           1 0.1977914 0.65983122
\end{Soutput}
\end{Schunk}
and the response \code{y}
\begin{Schunk}
\begin{Sinput}
R> head(bf$y)
\end{Sinput}
\begin{Soutput}
         num
1  0.2232725
2  0.2479576
3  0.1221580
4 -0.1370822
5 -0.1108988
6 -0.1011208
\end{Soutput}
\end{Schunk}
to setup the optimizer function with:
\begin{Schunk}
\begin{Sinput}
R> lm.opt <- function(x, y, ...)
+  {
+    ## Only univariate response.
+    y <- y[[1L]]
+  
+    ## For illustration this is easier to read.
+    X <- x$mu$model.matrix
+  
+    ## Estimate model parameters.
+    par <- drop(chol2inv(chol(crossprod(X))) %*% crossprod(X, y))
+  
+    ## Set parameter names.
+    names(par) <- paste0("mu.p.", colnames(X))
+  
+    ## Return estimated parameters and fitted values.
+    rval <- list(
+      "parameters" = par,
+      "fitted.values" = drop(X %*% par),
+      "edf" = length(par),
+      "sigma" = drop(sqrt(crossprod(y - X %*% par) / (length(y) - ncol(X))))
+    )
+  
+    ## Set edf within .GlobalEnv for the
+    ## loglik() function in the lm_bamlss() family.
+    .lm_bamlss.p <<- length(par)
+  
+    return(rval)
+  }
\end{Sinput}
\end{Schunk}
This optimizer function can already be used with the \fct{bamlss} wrapper function and
all extractor functions are readily available.
\begin{Schunk}
\begin{Sinput}
R> f <- num ~ x1 + poly(x2, 5) + poly(x3, 5)
R> b <- bamlss(f, data = d, family = "lm", optimizer = lm.opt, sampler = FALSE)
R> summary(b)
\end{Sinput}
\begin{Soutput}
Call:
bamlss(formula = f, family = "lm", data = d, optimizer = lm.opt, 
    sampler = FALSE)
---
Family: LM 
Link function: mu = identity
*---
Formula mu:
---
num ~ x1 + poly(x2, 5) + poly(x3, 5)
-
Parametric coefficients:
             parameters
(Intercept)       0.199
x1               -0.653
poly(x2, 5)1     -1.791
poly(x2, 5)2      2.014
poly(x2, 5)3      0.474
poly(x2, 5)4     -1.752
poly(x2, 5)5      0.743
poly(x3, 5)1     -0.303
poly(x3, 5)2      4.273
poly(x3, 5)3      0.050
poly(x3, 5)4      0.572
poly(x3, 5)5      0.144
---
Optimizer summary:
-
edf = 12 sigma = 0.2257
---
\end{Soutput}
\begin{Sinput}
R> nd <- data.frame("x2" = seq(0, 1, length = 100))
R> nd$p <- predict(b, newdata = nd, term = "x2")
\end{Sinput}
\end{Schunk}
Plot the estimated effect of variable \code{x2}.
\begin{Schunk}
\begin{Sinput}
R> plot2d(p ~ x2, data = nd)
\end{Sinput}
\end{Schunk}

The next step is to setup a full Bayesian MCMC sampling function. Fortunately, if we
assume multivariate normal priors for the regression coefficients and an inverse Gamma prior
for the variance, a Gibbs sampler with multivariate normal and inverse Gamma full conditionals
can be created. The MCMC algorithm consecutively samples for $t = 1, \ldots, T$ from the full
conditionals
$$
\boldsymbol{\beta}^{(t)} | \cdot \sim N\left(\boldsymbol{\mu}_{\boldsymbol{\beta}}^{(t - 1)}, 
  \boldsymbol{\Sigma}_{\boldsymbol{\beta}}^{(t - 1)}\right)
$$
and
$$ 
{\sigma^2}^{(t)} | \cdot \sim IG\left({a^{\prime}}^{(t - 1)}, {b^{\prime}}^{(t - 1)}\right),
$$
where $IG( \cdot )$ is the inverse Gamma distribution for sampling the variance parameter.
The covariance matrix for $\boldsymbol{\beta}$ is given by
$$
\boldsymbol{\Sigma}_{\boldsymbol{\beta}} = \left(\frac{1}{\sigma^2}\mathbf{X}^\top\mathbf{X} +
  \frac{1}{\sigma^2}\mathbf{M}^{-1}\right)^{-1}
$$
and the mean
$$
\boldsymbol{\mu}_{\boldsymbol{\beta}} = \boldsymbol{\Sigma}_{\boldsymbol{\beta}}
  \left(\frac{1}{\sigma^2}\mathbf{X}^\top\mathbf{y} +
  \frac{1}{\sigma^2}\mathbf{M}^{-1}\mathbf{m}\right),
$$
where $\mathbf{m}$ is the prior mean and $\mathbf{M}$ the prior covariance matrix.
Similarly, for $\sigma^2$ paramaters $a^{\prime}$ and $b^{\prime}$ are computed by
$$
a^{\prime} = a + \frac{n}{2} + \frac{p}{2}
$$
and
$$
b^{\prime} = b + \frac{1}{2}(\mathbf{y} -
  \mathbf{X}\boldsymbol{\beta})^\top(\mathbf{y} - \mathbf{X}\boldsymbol{\beta}) +
  \frac{1}{2} (\boldsymbol{\beta} - \mathbf{m})^\top \mathbf{M}^{-1}(\boldsymbol{\beta}
  - \mathbf{m}),
$$
where $a$ and $b$ are usually set small, e.g., with $a = 1$ and $b = 0.0001$, such that the prior is
flat and uninformative.

We can implement the MCMC algorithm in the following sampling function
\begin{Schunk}
\begin{Sinput}
R> lm.mcmc <- function(x, y, start = NULL,
+    n.iter = 12000, burnin = 2000, thin = 10,
+    m = 0, M = 1e+05,
+    a = 1, b = 1e-05,
+    verbose = TRUE, ...)
+  {
+    ## How many samples are saved?
+    itrthin <- seq.int(burnin, n.iter, by = thin)
+    nsaves <- length(itrthin)
+  
+    ## Only univariate response.
+    y <- y[[1L]]
+  
+    ## For illustration this is easier to read.
+    X <- x$mu$model.matrix
+  
+    ## Again, set edf within .GlobalEnv for the
+    ## loglik() function in the lm_bamlss() family.
+    .lm_bamlss.p <<- ncol(X)
+  
+    ## Number of observations and parameters.
+    n <- length(y)
+    p <- ncol(X)
+  
+    ## Matrix saving the samples.
+    samples <- matrix(0, nsaves, p + 1L)
+  
+    ## Stick to the naming convention.
+    pn <- paste0("mu.p.", colnames(X))
+    colnames(samples) <- c(
+      pn,      ## Regression coefficients and
+      "sigma"  ## variance samples.
+    )
+  
+    ## Setup coefficient vector,
+    ## again, use correct names.
+    beta <- rep(0, p)
+    names(beta) <- pn
+    sigma <- sd(y)
+  
+    ## Check for starting values obtained,
+    ## e.g., from lm.opt() from above.
+    if(!is.null(start)) {
+      sn <- names(start)
+      for(j in names(beta)) {
+        if(j %in% sn)
+          beta[j] <- start[j]
+      }
+    }
+  
+    ## Process prior information.
+    m <- rep(m, length.out = p)
+    if(length(M) < 2)
+      M <- rep(M, length.out = p)
+    if(!is.matrix(M))
+      M <- diag(M)
+    Mi <- solve(M)
+  
+    ## Precompute cross products.
+    XX <- crossprod(X)
+    Xy <- crossprod(X, y)
+  
+    ## Inverse gamma parameter.
+    a <- a + n / 2 + p / 2
+  
+    ## Start sampling.
+    ii <- 1
+    for(i in 1:n.iter) {
+      ## Sampling sigma
+      b2 <- b + 1 / 2 * t(y - X %*% beta) %*% (y - X %*% beta) +
+        1 / 2 * t(beta - m) %*% Mi %*% (beta - m)
+      sigma2 <- sqrt(1 / rgamma(1, a, b2))
+  
+      ## Sampling beta.
+      sigma2i <- 1 / sigma2
+      Sigma <- chol2inv(chol(sigma2i * XX + sigma2i * Mi))
+      mu <- Sigma %*% (sigma2i * Xy + sigma2i * Mi %*% m)
+      beta <- MASS::mvrnorm(1, mu, Sigma)
+        
+      if(i %in% itrthin) {
+        samples[ii, pn] <- beta
+        samples[ii, "sigma"] <- sqrt(sigma2)
+        ii <- ii + 1
+      }
+      if(verbose) {
+        if(i %% 1000 == 0)
+          cat("iteration:", i, "\n")
+      }
+    }
+  
+    ## Convert to "mcmc" object.
+    samples <- as.mcmc(samples)
+  
+    return(samples)
+  }
\end{Sinput}
\end{Schunk}
The new sampling function can be directly used with the \fct{bamlss} wrapper
\begin{Schunk}
\begin{Sinput}
R> b <- bamlss(f, data = d, family = "lm", optimizer = lm.opt, sampler = lm.mcmc)
\end{Sinput}
\begin{Soutput}
iteration: 1000 
iteration: 2000 
iteration: 3000 
iteration: 4000 
iteration: 5000 
iteration: 6000 
iteration: 7000 
iteration: 8000 
iteration: 9000 
iteration: 10000 
iteration: 11000 
iteration: 12000 
\end{Soutput}
\begin{Sinput}
R> summary(b)
\end{Sinput}
\begin{Soutput}
Call:
bamlss(formula = f, family = "lm", data = d, optimizer = lm.opt, 
    sampler = lm.mcmc)
---
Family: LM 
Link function: mu = identity
*---
Formula mu:
---
num ~ x1 + poly(x2, 5) + poly(x3, 5)
-
Parametric coefficients:
                 Mean     2.5%      50%    97.5% parameters
(Intercept)   0.19937  0.11443  0.19799  0.28402      0.199
x1           -0.65380 -0.79612 -0.65227 -0.51375     -0.653
poly(x2, 5)1 -1.79249 -2.71924 -1.79813 -0.86320     -1.791
poly(x2, 5)2  2.00010  1.08057  1.99022  2.88333      2.014
poly(x2, 5)3  0.48731 -0.48780  0.49488  1.49446      0.474
poly(x2, 5)4 -1.78055 -2.66888 -1.78927 -0.81921     -1.752
poly(x2, 5)5  0.73783 -0.28079  0.74257  1.72215      0.743
poly(x3, 5)1 -0.29363 -1.25928 -0.29305  0.56080     -0.303
poly(x3, 5)2  4.29066  3.36496  4.28125  5.26903      4.273
poly(x3, 5)3  0.03256 -0.98550  0.05819  1.01670      0.050
poly(x3, 5)4  0.57143 -0.36514  0.58563  1.48913      0.572
poly(x3, 5)5  0.13204 -0.81059  0.13434  1.07470      0.144
---
Sampler summary:
-
DIC = 23.3951 pd = 52.4174 runtime = 2.483
---
Optimizer summary:
-
edf = 12 sigma = 0.2257
---
\end{Soutput}
\begin{Sinput}
R> ## Predict for all terms including 95% credible intervals
R> nd$x1 <- nd$x3 <- seq(0, 1, length = 100)
R> for(j in c("x1", "x2", "x3")) {
+    nd[[paste0("p.", j)]] <- predict(b, newdata = nd, term = j,
+      FUN = c95, intercept = FALSE)
+  }
\end{Sinput}
\end{Schunk}
The estimated effects can be plotted with:
\begin{Schunk}
\begin{Sinput}
R> par(mfrow = c(1, 3))
R> plot2d(p.x1 ~ x1, data = nd, fill.select = c(0, 1, 0, 1), lty = c(2, 1, 2))
R> plot2d(p.x2 ~ x2, data = nd, fill.select = c(0, 1, 0, 1), lty = c(2, 1, 2))
R> plot2d(p.x3 ~ x3, data = nd, fill.select = c(0, 1, 0, 1), lty = c(2, 1, 2))
\end{Sinput}
\end{Schunk}

\end{appendix}

\end{document}